\documentclass[11pt,leqno,fleqn]{article}

\usepackage{lmodern}        
\usepackage{mathrsfs}
\usepackage{amsfonts}

\usepackage[margin=1in]{geometry}      
\usepackage{amsmath}                   
\usepackage{amssymb}                   
\usepackage{amsthm}
\usepackage{booktabs}
\usepackage{mathtools}                 
\usepackage{enumitem}                  
\usepackage{microtype}                 
\usepackage{parskip}                   
\usepackage{titlesec}                  

\usepackage{parskip}                   
\usepackage{setspace}    
\usepackage{xcolor}
\definecolor{blue}{RGB}{0, 51, 153}      
\usepackage{hyperref}
\hypersetup{
     colorlinks = false,       
     urlbordercolor=green,     
     linkcolor=blue,
     filecolor=magenta,      
     urlcolor=cyan, 
     linkbordercolor=red     
    }
\usepackage{tikz}            
\usepackage{tikz-cd}  
\usetikzlibrary{positioning, arrows.meta}

\usepackage[normalem]{ulem}    

\definecolor{Lightgreen}{rgb}{0.8,1,0.6} 
\definecolor{Lightgray}{gray}{.80}
\definecolor{Myblue}{rgb}{.8,0.85,1}         

\renewcommand\fbox{\fcolorbox{blue}{white}}
 
\newcommand{\HItxtG}[1]{{\fcolorbox{green}{white}{#1}}}

\newcommand{\Hide}[1]{}
\newcommand{\ie}{\textit{i.e.}}
\newcommand{\eg}{\textit{e.g.}}
\newcommand\QED{$\square$}

\newcommand{\size}[1]{\bigl|#1\bigr|}
\newcommand{\Set}[1]{\{ #1 \}}
\newcommand{\SET}[1]{\Bigl\{ #1 \Bigr\}}

\newcommand{\nats}{\mathbb{N}}

\newcommand{\ints}{\mathbb{Z}}

\newtheorem{theorem}{Theorem}

\newtheorem{fact}[theorem]{Fact}
\newtheorem{restxxx}[theorem]{Restriction}

\newtheorem{agreexxx}[theorem]{Agreement}

\newtheorem{termxxx}[theorem]{Terminology}

\newtheorem{notxxx}[theorem]{Notation}

\newtheorem{assumxxx}[theorem]{Assumption}

\newtheorem{convenxxx}[theorem]{Convention}

\newtheorem{exaxxx}[theorem]{Example}
\newenvironment{example}{\begin{exaxxx}\rm}%
     {\hfill\QED\end{exaxxx}}
\newtheorem{exaxxxsansQED}[theorem]{Example}     
\newenvironment{examplesansQED}{\begin{exaxxxsansQED}\rm}%
   {\end{exaxxxsansQED}}
\newtheorem{exaxxxTitle}[theorem]{Example}     
     {\end{exaxxxTitle}}  
\newtheorem{exexxx}[theorem]{Exercise}

\newtheorem{exexxxTitle}[theorem]{Exercise}     
     {\end{exexxxTitle}}
\newtheorem{remxxx}[theorem]{Remark}
\newenvironment{remark}{\begin{remxxx}\rm}{\end{remxxx}}
\newtheorem{openxxx}[theorem]{Open Problem}
\newenvironment{problem}{\begin{openxxx}\rm}{\end{openxxx}}%
\newtheorem{opensmallxxx}[theorem]{Problem}
\newtheorem{conjxxx}[theorem]{Conjecture}

\newtheorem{defxxx}[theorem]{Definition}
\newenvironment{definition}[1]{\begin{defxxx}[\emph{#1}]\rm}%
   {\hfill\QED\end{defxxx}}
\newtheorem{defxxxsansQED}[theorem]{Definition}
\newenvironment{definitionsansQED}[1]{\begin{defxxxsansQED}[\emph{#1}]\rm}%
   {\end{defxxxsansQED}}
\newtheorem{Prxxx}[theorem]{Proof}
{\end{Prxxx}} 

  {\addtolength{\leftskip}{#1}\addtolength{\rightskip}{#2}}{\par}

\title{\textsc{String Rewriting Systems}:\\[0.5ex]
       Brief Introduction and Sample of Open Problems}
\author{Assaf Kfoury}
\date{29 June 2026}

\begin{document}

\maketitle

This document is the result of pulling together small parts of
different lecture notes I have written over several decades for first-year 
graduate-level courses --- typically with the title \emph{Formal Methods} ---
which included many other topics of mathematical logic
and theoretical computer science.
After combining these materials, I also tried to update the references
and (minimally) adjust the text to account for progress accomplished in the intervening years.

The sample of open problems in 
\HItxtG{\color{blue}{Section~\ref{sect:list-of-problems}}} 
below is a small collection of special cases that still 
appear unresolved up until the date of this writing. 
There is nothing particularly special 
about my sample other than it includes
open problems that were scattered in my old lecture notes
from different years --- and that
are simple enough to state and present in lecture in connection with other topics
of model theory, or proof theory, or computability. 

    Moreover, my sample is \emph{not} a survey; it is a small fraction of much
    longer lists of open problems that researchers have accumulated over the years.
    See, for example, the 
    {\href{https://tlca.di.unito.it/opltlca/opltlca.pdf}{\color{blue}{TLCA List of Open Problems}}}
    or the {\href{https://www.cs.tau.ac.il/~nachum/rtaloop/problems/complete.pdf}{\color{blue}{RTA List of Open Problems}}}.
    Both of these lists include many problems on string rewriting or related to string rewriting, and
    both are far more comprehensive than my sample in this document, although they
    still omit many of the problems included in mine.

\newpage

{ \setstretch{1.65}
\color{blue}{
\tableofcontents
}
}

\newpage

\section{String Rewriting Systems: Preliminary Definitions}

A \emph{string rewriting system} (SRS), historically referred to as a \emph{semi-Thue system}, 
is a pair
\[
T=(\Sigma,R),
\]
where \(\Sigma\) is an alphabet and \(R\) is a set of rewrite rules,
each of the form:
\[
\ell \to r,
\]
with $\ell,r \in \Sigma^*$.  Here $\Sigma^*$ denotes the set of all finite \emph{words} or 
\emph{strings} (we use the two words interchangeably)
over $\Sigma$, including the empty word $\varepsilon$. We say 
the system $T$ is \emph{infinite} if $\Sigma$ or $R$ is an
infinite set, otherwise $T$ is a \emph{finite} SRS. We will restrict
attention to the case when $\Sigma$ is finite, but leave open the possibility that
$R$ may be an infinite set of rules.

The one-step rewriting relation induced by \(T\) is written $u \to_T v$, where $u,v\in\Sigma^*$, and is
defined by:
\begin{alignat*}{5}
    & u \to_T v \quad \Longleftrightarrow\quad\ \ 
       &&\text{there exist words $p,q,\ell,r \in \Sigma^*$ such that} \\
    & && u=p\,{\ell}\,q,\quad v=p\,r\,q, \quad\text{and}\quad \ell \to r \in R.   
\end{alignat*}
Thus, a rewrite step replaces an occurrence of the left-hand side \(\ell\) by the right-hand side \(r\), inside an arbitrary 
context $ p\,[\ \ ]\,q$. The square brackets ``$[$" and ``$]$" are not part of the syntax and are temporary delimeters of
an omitted substring. 

The \emph{reflexive and transitive closure} of $\to_T$ is written \(\to_T^*\),
so that \(u \to_T^* v\) means that \(v\) can be obtained from \(u\) by zero or more applications of rules from \(R\).

The \emph{reflexive, symmetric, and transitive closure} of $\to_T$ is written \(\leftrightarrow_T^*\),
so that \(u \leftrightarrow_T^* v\) means that \(v\) can be obtained from \(u\) by 
allowing rules from $R$ to be applied both \emph{forward} and \emph{backward}.  
The relation \(\leftrightarrow_T^*\) is an
equivalence relation, often denoted ${\approx}_T$ .%
   
   When rewriting is extended to the use of \(\leftrightarrow_T^*\) instead of \(\to_T^*\),
   the resulting system is sometimes called a \emph{Thue system}. Put differently, a \emph{Thue system} is a 
   special case of a SRS (\ie, a semi-Thue system) when, for every rule $\ell \to r \in R$, its
   reverse $r \to \ell$ is also included in $R$.   
   %

We may omit the subscript ``$T$" from ``$\to_T$", ``$\to_T^*$" and ``\(\leftrightarrow_T^*\)" --- and write instead
``$\to$", ``$\to^*$" and ``\(\leftrightarrow^*\)" ---
whenever it is clear that ``$\to$" refers to a one-step rewrite rule.

We will use early letters of the Latin alphabet, such as $a,b,c,\ldots$, 
to name individual symbols of $\Sigma$, and middle or late letters, such as
$\ell, r, u, v, w$, possibly with subscripts, to range over strings
in $\Sigma^*$. Early letters with numeral subscripts, such as $a_2,a_5,b_3,\ldots$,
may also be used as names for individual symbols of $\Sigma$. 
By contrast, early letters with non-numeral subscripts, such as 
$a_i, a_j, b_k,\ldots$, should be understood as ranging 
over symbols of $\Sigma$, not as fixed names of particular symbols in
$\Sigma$. 

Thus \emph{ranging letters}, unlike \emph{naming letters}, may denote equal 
objects: for example, we may have $u=v$, or $a_i=a_j$ because $i=j$. 
But if the letters $a$ and $b$ are names of distinct symbols of $\Sigma$, then 
always $a\neq b$ and also $a\neq \varepsilon \neq b$ where the letter $\varepsilon$ 
names a ``blank space" (the empty word).%
    \footnote{
    $\phantom{s}$ In a different context --- say, for example, first-order logic --- we will
    say that ``{ranging letters}" are \emph{metavariables}, \ie, variables outside
    the formal syntax of first-order logic, and ``naming letters" are \emph{constant symbols} inside the formal syntax of first-order logic.    
    }

We will use early letters of the Greek alphabet, such as
$\alpha, \beta, \gamma, \ldots$, to range over finite and infinite
rewrite sequences.

\begin{example}
Let $\Sigma=\{a,b,c\}$ and $R=\{a\,b\to b\,a, b\,c\to c\,b\}$. Then $a\,a\,b\,c \to a\,b\,a\,c$ because 
$a\,a\,b\,c = a\,[a\,b]\,c$ and the rule \(a\,b\to b\,a\) gives $a\,[a\,b]\,c \to a\,[b\,a]\,c = a\,b\,a\,c$. 
\textbf{Remember:} \emph{``$[$" and ``$]$" are not part of the syntax and we will use them repeatedly 
in examples to follow.} So that also, with two rewrite steps, we can produce the sequence:
\[
   a\,[a\,b]\,c \to\ [a\,b]\,a\,c\ \to b\,a\,a\,c\quad\text{or, more succintly,}\quad 
   a\,a\,b\,c \to^* b\,a\,a\,c .
\]
Since no rule can be applied to $b\,a\,a\,c$, we say that $b\,a\,a\,c$ is \emph{irreducible} and also
say that $b\,a\,a\,c$ is a \emph{normal form} of
$a\,a\,b\,c$, in which case we may write $b\,a\,a\,c\in \operatorname{NF}(a\,a\,b\,c)$. Note that $a\,a\,b\,c$
has more than one normal forms, because:
\[
  a\,a\,[b\,c] \to\ a\,a\,c\,b \quad \text{and\quad $a\,a\,c\,b$ is irreducible. }
\]
For the SRS under consideration, it is easy to check that $a\,a\,b\,c$
has exactly two normal forms, and we can write $\operatorname{NF}(a\,a\,b\,c) = \Set{b\,a\,a\,c,\ a\,a\,c\,b}$ .
We can combine the two reductions of $a\,a\,b\,c$ with the following diagram, with labeled brackets to distinguish
the two overlapping contexts:

\qquad \begin{tikzpicture}
    \node (leftnode) at (0,0) {$a\,\underset{\text{\tiny 1}}{[}a\,\underset{\text{\tiny 2}}{[} b \underset{\text{\tiny 1}}{]}\,c\underset{\text{\tiny 2}}{]}$};
    \node (leftnode2) at (0,.15) {$\phantom{SSSSSSS}$};
    
    \node (upnode) at (3.5,.56) {$[a\,b]\,a\,c$};
    \node (upnode2) at (3.5,.6) {$\phantom{SSSS}$};
    \node (upnode3) at (5.8,.56) {$b\,a\,a\,c$};
    \node (downnode) at (3.5,-.45) {$a\,a\,c\,b$};
    \node (downnode2) at (3.5,-.6) {$\phantom{SSSS}$};
    
    \draw[->] (leftnode2) -- node[above=2pt] {\scriptsize $1$} (upnode2);
    \draw[->] (upnode) -- (upnode3);
    
    \draw[->] (leftnode2) -- node[below=2pt] {\scriptsize $2$} (downnode2);
\end{tikzpicture}

     The pair $\langle a\,b\,a\,c\,,\,a\,a\,c\,b\rangle$ is called a \emph{critical pair},
     the result of applying two rules 
     (or two instances of the same rule) to overlapping substrings in the string $a\,a\,b\,c $, 
     in two different competing ways.

If we extend the set $R$ of rules to another set $R'$ which includes both a rule and its reverse: 
\[
    R'\ =\ \{a\,b\to b\,a,\ b\,a\to a\,b,\ b\,c\to c\,b,\ c\,b\to b\,c\}
\]
the resulting SRS is a Thue, not only semi-Thue, system.
\end{example}

\begin{example}
    This is an example of an infinite SRS 
    $T = (\Sigma,R)$ where $\Sigma=\{a,b\}$ with an infinite set $R$ of rules:
    \[
    R\ =\ \Set{b \to a\,a}\ \cup\ \Set{ a^n\,b \to b\,a^{n}\;|\; n\geqslant 1\,}
    \quad\text{where $a^n$ stands for } \underbrace{a\,a\,\cdots\,a}_{n\;\text{times}} .
    \]
     In this SRS, every string containing only $a$'s is irreducible, and every string
     containing $m$ occurrences of $a$ and $n$ occurrences of $b$ has a unique normal form
     $a^{m+2n}$. Rewriting in this SRS always terminates at a unique normal form.
     The system will be more complicated if the rewrite rules are:
    \[
    R_1\ =\ \Set{b \to a\,a}\ \cup\ \Set{ a^n\,b \to b\,a^{n+1}\;|\; n\geqslant 1\,} .
    \]
    In the resulting SRS  $T_1 = (\Sigma,R_1)$ every string in $\Sigma^*$ has a normal form,
    which may or may not be unique. For example, $\operatorname{NF}(a\,a\,b) = \Set{a^5\,,\,a^6}$  
    because $a\,a\,b \to^* a^5$ and $a\,a\,b \to^* a^6$ in $T_1$.
\end{example}
 
\section{String Rewriting Systems versus
         Term Rewriting Systems}

A \emph{string rewriting system} $T=(\Sigma,R)$ is not a \emph{term rewriting system} (TRS). However, it can be represented very naturally as a TRS as we explain next.

Recalling from other parts of model theory and proof theory, 
a first-order signature consists of a set of variables, a set of constant symbols, 
and a set of function symbols (each with an arity $\geqslant 1$). A \emph{first-order term},
or simply a \emph{term}, is then inductively defined by the following three rules:
\begin{enumerate}
    \item 
        \textbf{Variables}: Every variable symbol $x, y, z, \ldots$ is a term.  
    \item 
        \textbf{Constants}: Every constant symbol $c, d, \ldots$ is a term.
    \item  
        \textbf{Functions}: If $f$ is an $n$-ary function symbol (where $n \geqslant 1$) and 
          $t_1,t_2,\ldots,t_n$ are already defined terms, then the expression 
          $f(t_1,t_2,\ldots,t_n)$ is a term.
\end{enumerate}

Reverting back to our SRS $T=(\Sigma,R)$, we can take a word
\[
a_1\,a_2\,\cdots\, a_n \quad\text{where $a_1,\ldots,a_n \in \Sigma$},
\]
as the unary first-order term
\[
a_1\, (a_2\, (\cdots \, a_n\, (\varepsilon)\,\cdots)).
\]
where we treat every alphabet symbol $a_i\in\Sigma$ as a unary function symbol and \(\varepsilon\) as a constant.
For example, the word $a\,b\,c$ is represented as the term $a\,(b\,(c\, (\varepsilon)))$.

A rewrite rule such as $a\,b\to b\,a$ becomes the first-order rewrite rule
$a\,(b\,(x)) \to b\,(a\,(x))$, where the $x$
represents the remaining suffix of the word.
More generally, a string-rewrite rule of the form:
\[
a_1\,a_2\,\cdots\, a_m\ \to\ b_1\,b_2\,\cdots b_n
\]
is translated to a (first-order) term rewrite rule: 
\[
a_1\,(a_2\,(\cdots\, a_m\, (x)\,\cdots))
\ \to
\ b_1\,(b_2\,(\cdots\, b_n\,(x)\,\cdots)).
\]
This is a genuine first-order rewrite rule.  There is no binding, no abstraction, and no substitution of functions for functions.  Variables range only over first-order terms, here representing word suffixes.

\begin{remark}
    The first time one encounters rewrite rules, it may be a little confusing when comparing the ways in which
    they are used in \emph{string-rewriting systems} and in \emph{term-rewriting systems}. 

    In a \emph{string-rewriting system}, the rules operate on strings, \ie, ordered sequences of characters. 
    Every symbol in a rule is a literal, concrete character from the alphabet $\Sigma$. There is no concept 
    of a ``placeholder" or ``variable." If a string-rewriting system includes the rule 
    $a\,b \to b\,a$, it means you can replace the literal string ``$a\,b$" with ``$b\,a$"; it does not
    mean you can match its left-hand side with $c\,b$, or $d\,b$, or anything else. The match must be exact, character-for-character.

    In contrast, a \emph{term-rewriting system} operates on first-order terms. Because terms contain variables, 
    the rewrite rules themselves can contain variables. Instead of a literal string substitution, a rule in a  
    term-rewriting system acts as a structural pattern.
    When a rule is applied, the variables in the rule can be instantiated (substituted) with any valid term.
    For example, $x \cdot (y + z) \rightarrow (x \cdot y) + (7\cdot x \cdot z)$ as a term-rewrite rule involves two
    binary function symbols $\Set{+,\cdot}$, three variables $\Set{x,y,z}$, and one
    constant symbol $\Set{7}$, which we can use to carry out the following rewrite
    $5 \cdot ((3 + a) + b) \rightarrow (5 \cdot (3 + a)) + (7\cdot 5 \cdot b)$ by making three substitutions
    $\Set{x\mapsto 5,\ y\mapsto (3+a),\ z\mapsto b}$.
    \hfill \QED
\end{remark}

\section{Computational Universality}
\label{sect:universality}

String-rewriting systems are computationally universal. They can simulate the operation of a Turing machine by 
encoding configurations as words and transitions as string-rewrite rules. Consequently, many natural 
decision problems for SRSs are undecidable in full generality. Here, we present some of the most 
frequently cited ones, which we can limit to the case of finite SRSs.
 
There is no algorithm 
which, when given an arbitrary finite string-rewriting 
system $T = (\Sigma,R)$ and words \(u,v\in\Sigma^*\), can decide any of the following problems:

\begin{enumerate}[label=(\arabic*)]
\item \textbf{Reachability:} whether $u\to_T^* v$.

\item \textbf{Word Problem:} 
whether $u \ {\approx}_T\ v$, \ie,\ whether \(u \leftrightarrow_T^* v\). 

\item \textbf{Common Descendant:} whether there exists a word $w$ such that $u\to_T^* w$ and $v\to_T^* w$.
\end{enumerate}

Separate from the preceding three, the following is another foundational result. 
There is no algorithm 
which, when given an arbitrary finite string-rewriting system $T = (\Sigma,R)$, can decide:
\begin{enumerate}[resume, label=(\arabic*)]
    \item \textbf{Termination:} whether, for every \(u\in\Sigma^*\), every rewrite sequence from $u$ terminates,\\[1.5ex]
    or, stated differently, whether there is a word \(w_0\in\Sigma^*\) and there is an infinite rewrite sequence from $w_0$
    such that \(w_0\to_T w_1\to_T w_2\to_T\cdots\) \ .
\end{enumerate}

  The termination problem is sometimes called \textbf{\emph{uniform termination}} or \textbf{\emph{global termination}},
  in order to distinguish it from the \emph{local termination} problem (below). If we do not qualify termination
  with the adjective ``uniform" or ``local", then we mean \textbf{\emph{global termination}}.

  There is no algorithm which, when given an arbitrary finite string-rewriting system $T = (\Sigma,R)$ together with  
  a word \(u\in\Sigma^*\), can decide:
  \begin{enumerate}[resume, label=(\arabic*)]
     \item 
     \textbf{Local Termination:}  whether every rewrite sequence starting from $u$ terminates.
  \end{enumerate}
  
The undecidability of these problems is not surprising, once we know that SRSs
can simulate arbitrary computations by Turing machines.
For later considerations, we note that, although:
\begin{itemize}
    \item 
    \emph{SRS $T$ is globally terminating} $\Longleftrightarrow$ 
    \emph{$T$ is locally terminating on every $u\in {\Sigma}^*$},
\end{itemize}
it is still possible that for a class of SRSs, \textbf{\emph{local termination}} is decidable while 
\textbf{\emph{global termination}} is undecidable.%
    \footnote{
    $\phantom{s}$ Put differently, the unsolvability degree of \textbf{\emph{global termination}} is higher
    than the unsolvability degree of \textbf{\emph{local termination}}. This is similar to a difference
    between the \emph{Totality Problem} and the \emph{Halting Problem} for Turing machines:
    the former's unsolvability degree (${\Pi}^0_2$-complete) is higher than
    the latter's unsolvability degree (${\Sigma}^0_1$-complete).
    }
We give an example of such a class of SRSs in 
\HItxtG{\color{blue}{Subsection~\ref{subsect:length-preserving}}}.

\section{Natural Decidable Restrictions}

There is no single, canonical, ``least restrictive'' condition that makes important problems of SRSs decidable. Different restrictions make different problems decidable,
of which we can then study the computational complexity.
The following are among natural and important restrictions.

\subsection{Length-Preserving Systems and Length-Reducing Systems}
\label{subsection:length} 

A string-rewriting system $T = (\Sigma,R)$ is \emph{length-preserving}, resp. \emph{length-reducing}, if
\[
   \text{every rule}\ \ell \to r \in R\quad \text{satisfies}\quad |\ell|=|r|,\ \text{resp.}\ |\ell|>|r| .
\]
For fixed input words \(u\) and \(v\), if all the rules are length-preserving, 
then any word reachable from \(u\) has the same length as \(u\).  
Since there are only finitely many words of a fixed length over a finite alphabet, 
reachability from \(u\) can be decided by a finite exhaustive search.
Thus, for length-preserving systems, the problem $u\to_T^* v?$ is decidable.

Similarly, if all the rules are length-reducing, then any word reachable from \(u\) has length at most \(|u|\).  Hence,
reachability from \(u\) is again decidable by finite search.

On the other hand, while a length-preserving SRS may allow infinite
rewrite sequences, a length-reducing SRS is always 
terminating, since in the latter, every rewrite step decreases
the length of the word. An example of a length-preserving SRS that 
allows infinite rewrite sequences is the following:
$T = (\Set{a,b,c}\ ,\ \Set{a\to c,\ b\to c,\ a\,b\to b\,a,\ b\,a\to a\,b})$.

\subsection{Terminating Systems}
\label{subsect:terminating}

A string-rewriting system $T = (\Sigma,R)$ is \emph{terminating}, or \emph{Noetherian}, if there is no infinite sequence
\[
w_0\to_T w_1\to_T w_2\to_T\cdots.
\]

If a \emph{finite} SRS is known to be terminating on every word, then from any fixed starting word $u$, 
only finitely many words are reachable.  This is because the rewrite graph is finitely branching, 
since the SRS is finite, and an infinite reachable tree would contain an infinite path by K\"{o}nig's Lemma.%
    \footnote{
    $\phantom{s}$ By K\"{o}nig's Lemma, every 
    finitely branching infinite tree must have an infinite path. Further details on this
    fact are in the {\href{https://en.wikipedia.org/wiki/Konig's_lemma}{\color{blue}{Wikipedia page}}}.
    }
Hence, \emph{given a terminating \emph{finite} string-rewriting system $T$, the reachability problem in $T$ 
is decidable.} 

However, there is an important caveat: termination itself is undecidable in general,
as indicated in \HItxtG{\color{blue}{Section~\ref{sect:universality}}}.  
Moreover, the restriction to \emph{finite} SRSs is crucial, as demonstrated by the following example
of an \emph{infinite} SRS which is terminating but whose reachibility problem is undecidable.

\begin{example}
    Let $S \subseteq \nats$ be an infinite undecidable set of natural numbers, and
    consider the following \emph{infinite} string-rewriting system
    $T = \big(\Set{a, b, c}\ ,\ \Set{b\to c}\cup\Set{a\to b^n\;|\; n\in S}\big)$.

    Consider what happens if we start with the one-character word $ w = a $. 
    Is it terminating? Yes! From $a$, you can jump to $b^n$ in one step. From $b^n$, you can replace each 
    $b$ with a $c$, step-by-step until you reach $c^n$. From $c^n$, no more rules apply. Every single rewrite
    sequence is guaranteed to stop in at most $n+1$ steps. There are no loops and no infinite paths.

    But is reachability decidable? No! If you ask, ``Is the word $b^5$ reachable from $a$?", the answer is 
    \emph{yes} if and only if $5\in S$. Since we chose $S$ to be an undecidable set, there is no algorithm
    that can check whether the rule $a\to b^5$ is allowed by the system $T$.
\end{example}

\subsection{Terminating Length-Preserving Systems}
\label{subsect:length-preserving}

A common point of confusion in string rewriting arises when analyzing the boundary between 
\emph{local termination} and \emph{global termination} of length-preserving string-rewriting systems. 
If an undecidable problem like the \emph{Post Correspondence Problem} (PCP) or the Turing machine 
\emph{Halting Problem} can be encoded into a length-preserving tring-rewriting system, it appears to create 
a logical paradox if one assumes global termination of the latter is decidable. The resolution lies 
entirely in the strict decoupling of the two termination problems.

\begin{itemize}
    \item 
    \textbf{local termination} for length-preserving SRSs is \emph{decidable}.
    \item 
    \textbf{global termination} for length-preserving SRSs is \emph{undecidable}.
\end{itemize}

Among other methods to prove the latter undecidability result is to encode the Halting Problem or the PCP
into a length-preserving string-rewriting system.%
  \footnote{
  $\phantom{s}$
  The first proof that global termination is undecidable for \emph{length-preserving SRSs} was established
  by A.~Caron \cite{caron1991}. The reduction utilizes the fact that an input string of a 
  length-preserving system can act as the 
  fixed-size tape of a Linear Bounded Automaton (LBA). Because the uniform halting problem for LBAs is undecidable 
  (\ie, determining whether an LBA halts on \emph{all} possible inputs), the corresponding global termination problem
   for length-preserving SRSs inherits this undecidability. 
   Subsequent literature has streamlined this proof by bypassing the intermediate LBA stage and reducing the
   Post Correspondence Problem directly to length-preserving string-rewriting systems \cite{sakai-wang1999}.  
  }

\subsection{Complete (also called Convergent) Systems}
\label{subsect:complete}

A  string-rewriting system $T$ is \emph{confluent} if, whenever $u\to_T^* v_1$ and $u\to_T^* v_2$,
there exists a word \(w\) such that $v_1\to_T^* w$ {and} $v_2\to_T^* w$, which is graphically depicted by a
commutative diagram:

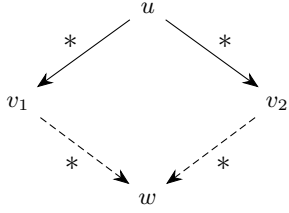
\qquad\quad
{\footnotesize 
\begin{tikzcd}[arrows=-{Stealth[scale=1.3]}, row sep=large, column sep=large, labels={font=\normalsize}]
& u \arrow[dl, "*"'] \arrow[dr, "*"] & \\
  v_1 \arrow[dr, dashed, "*"'] & & v_2 \arrow[dl, dashed, "*"] \\
& w & 
\end{tikzcd}
}\\[1.5ex]
A dashed arrow asserts the \emph{existence} of a rewriting sequence, so that the diagram is read as follows:
\textbf{for all} $u\to^* v_1$ and $u\to^* v_2$, \textbf{there exists}
a word $w$ such that $v_1\to^* w$ {and} $v_2\to^* w$.
Confluence is sometimes called the \emph{Church-Rosser property}.

A  string-rewriting system $T$ is \emph{complete}, or \emph{convergent}, if it is both \emph{terminating} and \emph{confluent}.
Convergence and completeness are often used synonymously in reports on SRSs. For simplicity here,
we will only use the words ``complete" and ``completeness".

In a complete SRS, every word \(u\) has a unique normal form, written \(\operatorname{nf}(u)\).
Put differently, the set $\operatorname{NF}(u)$ contains exactly one member, which is  $\operatorname{nf}(u)$, \ie,
\(\size{\operatorname{NF}(u)} = 1\) and \(\operatorname{NF}(u) = \Set{\operatorname{nf}(u)}\).
For a complete SRS, the word problem for the induced congruence is decidable:
\[
u \approx_T v \quad \Longleftrightarrow \quad \operatorname{nf}(u)=\operatorname{nf}(v).
\]

\begin{example}
  The following string-rewriting system $T = (\Set{a,b},\ \Set{a\,b\to b\,a,\ b\,a\to a\,b})$  is confluent
  but not terminating. A non-terminating sequence is:
  \[
    a\,b\ \to_T\ b\,a\ \to_T\ a\,b \ \to_T\ b\,a\ \to_T \ \cdots
  \]
  A concrete instance of its confluence is on the word $a\,b\,a$:
  
\qquad\quad
{\footnotesize 
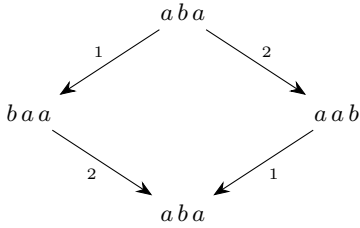
\begin{tikzcd}[arrows=-{Stealth[scale=1.3]}, row sep=large, column sep=large]
& a\,b\,a \arrow[dl, "1"'] \arrow[dr, "2"] & \\
b\,a\,a \arrow[dr, "2"'] & & a\,a\,b \arrow[dl, "1"] \\
& a\,b\,a & 
\end{tikzcd}
}\\[1.5ex]
where ``$1$" is the rule $a\,b\to b\,a$ and ``$2$" is the rule $b\,a\to a\,b$. The confluence on $a\,b\,a$
can be extended to any word $w\in \Set{a,b}^*$ containing occurrences of both $a$ and $b$.
\end{example}

\begin{example}
    The following string-rewriting system $T = (\Set{a,b,c,d},\ \Set{a\,b\to c,\ b\,c\to d})$  is terminating
    but not confluent. It is clearly terminating since it is length-reducing, but 
    it fails confluence because its two distinct rules compete for the same middle symbol, 
    splitting a single word into two irreconcilable paths.
    
    For a specific illustration of non-confluence by $T$, consider the word $a\,b\,c$, and denote by ``$1$"
    the rule $a\,b\to c$ and by ``$2$" the rule $b\,c\to d$. The word $a\,b\,c$ contains an overlapping critical pair
    which forces a permanent divergence:

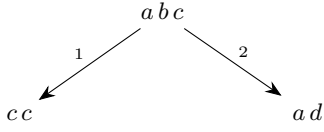
\qquad\quad
{\footnotesize 
\begin{tikzcd}[arrows=-{Stealth[scale=1.3]}, row sep=large, column sep=large]
& a\,b\,c \arrow[dl, "1"'] \arrow[dr, "2"] & \\
c\,c & & a\,d
\end{tikzcd}
}
 
Since $c\,c \neq a\,d$ and both are trapped at dead ends, they can never converge to a common descendant.
Thus, the system is not confluent.
\end{example} 

\subsection{Monadic Systems}

A string-rewriting system $T = (\Sigma,R)$ is often called \emph{monadic} if 
the right-hand side of every rule $\ell\to r\in R$ has length at most one, \(|r|\leqslant 1\).
Many treatments also assume that the left-hand side is at least as long as the 
right-hand side: \(|\ell|\geqslant |r|\) .

Monadic systems collapse words into single symbols or the empty word.  
They occur often in the study of formal languages and monoid presentations.
Combined with additional properties, especially confluence, monadic systems form well-studied decidable classes.  
For example, finite monadic confluent systems have particularly good decision properties for the corresponding word problems.

The next example shows that monadic and length-reducing are different ideas, although they often appear together.

\begin{example}
    The following string-rewriting system $T = (\Sigma,R)$ is both monadic and length-reducing,
    where $\Sigma=\Set{a,b,c}$ and $ R=\Set{a\,b\to c,\ b\,a\to c,\ c\,c\to c}$.
    Because it is length-reducing, $T$ is terminating.

    Extending the set of rules to $R_1 = R \cup \Set{c\,c\,c \to c\,c}$, the resulting system $T_1 = (\Sigma, R_1)$
    is no longer monadic, but still length-reducing and terminating.

    Extending the set of rules to $R_2 = R \cup \Set{c \to c}$, the resulting system $T_2 = (\Sigma, R_2)$
    is still monadic, but no longer length-reducing and terminating.

    None of the three systems ($T$, $T_1$ and $T_2$) is confluent. A counterexample is the word $w = a\,b\,a$:
    in all three systems, it holds that $w \to c\,a$ and $w \to a\,c$, where $c\,a$ and $a\,c$ are irreducible.
    Confluence can be enforced in the three systems by adding two rules, $a\to c$ and $b\to c$.
\end{example}
 
\section{Sample of Open Problems} 
\label{sect:list-of-problems}
 
This section lists several open problems, research directions, and nearby resolved results. Some of these
open problems are special cases of other open problems, in that solutions to the latter
imply solutions to the former. But I don't try to spell out the (mostly obvious) connections between
these open problems. I list them in an approximate order from \emph{less specific}
to \emph{more specific}, with the latter presuming some prior familiarity 
with \emph{group theory}, or \emph{monoid theory}, or some other area of \emph{abstract algebra}, or \emph{computability}.

\subsection{Termination for One-Rule String-Rewriting Systems}

One of the cleanest and best-known open problems is the following.

\medskip

\fbox{\parbox{0.985\linewidth}{
\begin{problem}
    \emph{
    Is there an algorithm which, when given an arbitrary one-rule string-rewriting system
    $T = (\Sigma, \Set{\ell \to r})$ where $\ell,r\in \Sigma^*$, 
    can decide whether the system is terminating?}
\end{problem}
}}

\medskip

For arbitrary finite SRSs, termination is undecidable, as indicated
in \HItxtG{\color{blue}{Subsection~\ref{subsect:terminating}}}. 
What is striking is that the case of single-rule SRSs in general is still not fully understood.   
There are simple cases of single-rule SRSs for which the problem is easily answered:
\begin{itemize}
    \item For a single \emph{length-reducing} rule ($|\ell| > |r|$), termination is trivially 
    guaranteed because the string gets shorter every time.
    \item For a single \emph{length-preserving} rule ($|\ell| = |r|$), termination is also easily 
    decidable because the system can only shuffle symbols within a fixed length.
\end{itemize}
 
When the length of words in a rewrite sequence can increase, there are partial positive results,
which we now describe. 
Termination is decidable for the class of one-rule SRSs that are restricted to \emph{grid}
rules or to be \emph{single-threaded}.

\begin{definition}{Grid Rule}
\label{def:grid}
      A rewrite rule $\ell \to r$ is called a \emph{grid rule} if there exists at least one letter $x$ in the alphabet
      $\Sigma$ such that $x$ occurs as often or less often on the right-hand side $r$ as it does on the left-hand 
      side $\ell$, in symbols, $\size{\ell}_x \geqslant \size{r}_x$ where $\size{u}_x$ 
      denotes the number of occurrences of the character $x$ in a string $u\in {\Sigma}^*$.
\end{definition}

\begin{examplesansQED}
    The following are \emph{grid rules}:
    \begin{enumerate}[label=(\arabic*)]
        \item 
        $a\,b\to b\,a$ is a grid rule. In fact, it is a \emph{strict grid} rule for both $a$ and $b$ 
        because $\size{a\,b}_a = \size{b\,a}_a = 1$ and $\size{a\,b}_b = \size{b\,a}_b = 1$ .
        \item 
        $a\,a\,b \to b\,b\,b\,b$ is a grid rule. Even though the string length increases, the symbol $a$
        appears twice on the left but zero times on the right. 
        \item 
        $b\to\varepsilon$ is a grid rule because the occurrence of $b$ decreases from $1$ to $0$.
    \end{enumerate}
    The following are \emph{not grid rules}:
    \begin{enumerate}[resume, label=(\arabic*)]
        \item 
        $a\to a\,a$ is not a grid rule. The only symbol present $a$ increases in count from $1$ to $2$.
        \item 
        $a\,b\to a\,a\,b\,b\,b$ is not a grid rule. 
        The count of $a$ increases from $1$ to $2$, and the count of $b$ increases from $1$ to $3$.
        \hfill \QED
    \end{enumerate}
\end{examplesansQED}

\bigskip
While frequently studied together, a string-rewriting system can be 
\emph{single-threaded} without being restricted to \emph{one rule}, and vice-versa.

\begin{definition}{Single-Threaded SRS}
 \label{def:single-threaded}  
 A string-rewriting system $T = (\Sigma,R)$ is \emph{single-threaded} if for every word $w\in {\Sigma}^*$, 
 there is \emph{at most one} $w'\in {\Sigma}^*$ such that $w \to_T w'$.  
\end{definition}

In a standard SRS, if multiple rules apply to a given word, or if a single rule matches at multiple positions,
the system exhibits non-deterministic branching. In a single-threaded SRS, branching pathways are eliminated
from the rewrite graph of every word, collapsing the derivation history into a linear chain or ``thread."

A string-rewriting system typically achieves the single-threaded property through one of two mechanisms:
\begin{itemize} 
    \item \textbf{Deterministic Rewriting Strategies:} 
    By enforcing a strict evaluation strategy, 
    such as \emph{leftmost} or \emph{rightmost} rewriting, the system is forced to behave deterministically. 
    Under a leftmost strategy, we scan the string from left to right and execute only the very 
    first match we encounter, ignoring any subsequent or overlapping matches further down the string.
    \item \textbf{Inherent System Geometry:} 
    A system can be naturally single-threaded if its entire set of rewrite rules is structurally conflict-free. 
    This occurs when no two rules can overlap with each other, and no single rule can overlap with itself 
    (\ie, the system is overlap-free).
\end{itemize} 

When a one-rule SRS $T = (\Sigma,\Set{\ell \rightarrow r})$ is forced to operate under a 
deterministic, single-threaded execution strategy (such as \emph{leftmost} rewriting), its global termination 
problem becomes {decidable} and its derivational complexity classes have been fully 
determined \cite{geser-hofbauer-waldmann2004}.%
   \footnote{
   $\phantom{s}$ That was the definitive paper around two decades ago by A. Geser, D. Hofbauer, and J. Waldmann 
   \cite{geser-hofbauer-waldmann2004}, in which they established this decidability result and mapped out the 
   derivational complexities for single-threaded one-rule string rewriting. It may be that there has been
   further progress on the termination problem for one-rule SRSs in the intervening years.}

\medskip

\fbox{\parbox{0.985\linewidth}{
\begin{problem}
    \emph{ Define other natural classes of one-rule SRSs for which termination, global or local,
    is decidable. }
\end{problem}
    }}

\subsection{The Boundary Between Decidability and Undecidability for Few Rules}

Another natural family of questions asks how many rewrite rules are needed 
before standard decision problems become undecidable.

Matiyasevich and S\'enizergues \cite{matiyasevich2005} proved that several important decision problems 
are already undecidable for string-rewriting systems with only three rules.  These include \emph{termination}, \emph{local termination}, \emph{reachability}, and \emph{common-descendant} problems, as defined in 
\HItxtG{\color{blue}{Section~\ref{sect:universality}}}.

Thus, the remaining uncertainty is concentrated in the very small cases.  
For example, the one-rule and two-rule cases are much more delicate than the general finite case.

\medskip

\fbox{\parbox{0.985\linewidth}{
\begin{problem}
    \emph{Determine the exact decidability boundary, in terms of the number of rules, for \emph{termination}, 
    \emph{reachability}, and \emph{related reachability} problems for string-rewriting systems.}
\end{problem} 
}}

\subsection{Reachability Languages for One-Rule String-Rewriting Systems}

For a fixed string-rewriting system \(T = (\Sigma,R)\) and a fixed word \(w\in\Sigma^*\), one can consider the set
\[
S_T(w)=\Set{\,w'\in\Sigma^* \;|\; w\to_T^* w'\,}.
\]
This is the set of all words reachable from \(w\).

For arbitrary SRSs, such sets can be highly complicated.  
For one-rule SRSs, however, the situation is subtler.  
Some special one-rule systems have reachable sets with good formal-language properties.  
For example, in some classes of one-rule grid SRSs, the reachable set \(S_T(w)\) 
is \emph{constructible context-free}, and certain sets of looping words are \emph{regular}.

This leads to several natural questions.

\medskip

\fbox{\parbox{0.985\linewidth}{
\begin{problem}
    \emph{For which classes of one-rule SRSs $T = (\Sigma\ ,\ \Set{\ell\to r})$ is the set of words
\(
S_T(w)=\Set{\,w'\in\Sigma^* \;|\; w\to_T^* w'\,}
\)
\emph{regular}, \emph{context-free}, \emph{effectively context-free}, or \emph{decidable} with useful complexity bounds?}%
\end{problem}
}}

\medskip

     Standard definitions of \emph{regular language}, \emph{context-free language}, and other
     formal languages, can be found at the
     {\href{https://en.wikipedia.org/wiki/Formal_language}{\color{blue}{Wikipedia page}}}, which also
     includes many references for further study.

\medskip

\subsection{Normalizing but Non-Terminating String-Rewriting Systems}

A word \(w\) is \emph{weakly normalizing} if there exists at least one reduction from \(w\) to a normal form.  
It is \emph{strongly normalizing} if all reductions from \(w\) terminate.

Recalling that ``system $T\ \text{is terminating} \Longleftrightarrow
\text{for all}\ w\in\Sigma^*,\; w \text{ is strongly normalizing}$," 
system $T$ may fail to be terminating, but still have many, or perhaps all, words admit some reduction (\ie, a rewrite sequence) to a normal form.
This leads to the following broad research direction.

\medskip

\medskip

\fbox{\parbox{0.985\linewidth}{
\begin{problem}
    \emph{Classify string-rewriting systems --- or characterize subclasses
    of them --- which are normalizing but not terminating.
    }
\end{problem}  
}}

\medskip

One version asks for conditions under which every word \(w\in\Sigma^*\) has at least one rewrite sequence
$w\to_T^* u$ to a normal form $u$, even though some words also admit infinite rewrite sequences.
This is the analogue in string-rewriting systems of the distinction between \emph{weak normalization}
and \emph{strong normalization} in the lambda calculus.

\subsection{Non-Terminating and Non-Looping One-Rule String-Rewriting Systems}

As indicated in \HItxtG{\color{blue}{Subsection~\ref{subsect:terminating}}}, 
a string-rewriting system $T = (\Sigma,R)$ is \emph{non-terminating}
if it admits an infinite reduction (\ie, an infinite rewrite sequence). 
A common source of non-termination is a \emph{loop}.  Informally, a loop occurs when a non-empty
word $w\in {\Sigma}^+$ rewrites, in one or more steps, to a word containing a copy of $w$ in context.%
   \footnote{
   $\phantom{s}$
   ${\Sigma}^+$ is the set of non-empty words over the alphabet $\Sigma$, \ie,
   ${\Sigma}^+ = {\Sigma}^* - \Set{\varepsilon}$.
   }
A \emph{loop} is a rewrite sequence of the form:
\[
w\ \to_T^+\ p\,w\, q\qquad \text{for some $p,q\in\Sigma^*$ } ,
\]
where $\to_T^+$ denotes one or more rewrite steps.
A SRS that admits a looping sequence is necessarily non-terminating. 
The converse is false in general: there are non-terminating 
SRSs that do not admit looping sequences.  The one-rule case, however,  
is elusive and still open to a solution.

\medskip

\fbox{\parbox{0.985\linewidth}{
\begin{problem}
    \emph{
    Does there exist a one-rule SRS $T = (\Sigma\ ,\ \Set{\ell \to r})$
    that is both non-terminating and non-looping?
    }
\end{problem}
}}

\medskip

\begin{example}
    To illustrate the subtle distinction between basic non-termination and looping behavior, we consider
    a three-rule SRS which is \emph{both non-terminating and non-looping}. 
    This system is a simplification of others considered by A.~Geser and H.~Zantema \cite{geser-zentama}.%
    \footnote{
     $\phantom{s}$ Geser and Zantema prove a stronger result; namely, there
     are non-terminating, non-looping SRSs having only \emph{two rules}. Their examples are however
     a little more complicated than the example we present here.
    }

    Let $T = (\Sigma, R)$ be a finite SRS over the alphabet $\Sigma = \Set{a, b, d}$ with
    the following three rewrite rules $R = \Set{b\,a \to d\,a,\ b\,d \to d\,b,\ a\,d \to a\,b\,b }$. 

   To demonstrate non-termination, we exhibit a single infinite rewrite sequence $\alpha$ that starts
   from the word $a\,b\,a$, which we break into consecutive, easy to inspect, fragments:
   \begin{align*}
    & a\,b\,a = a\,[b\,a] \xrightarrow{\;ba\to da\;} [a\,d]\,a \xrightarrow{\;ad\to abb\;} a\,b\,b\,a \\
    & a\,b\,b\,a = a\,b\,[b\,a] \xrightarrow{\;ba\to da\;} a\,[b\,d]\,a \xrightarrow{\;bd\to db\;} 
         [a\,d]\,b\,a \xrightarrow{\;ad\to abb\;} a\,b\,b\,b\,a \\
    & a\,b\,b\,b\,a = a\,b\,b\,[b\,a] \xrightarrow{\;ba\to da\;} a\,b\,[b\,d]\,a 
       \xrightarrow{\;bd\to db\;} a\,[b\,d]\,b\,a \xrightarrow{\;bd\to db\;} [a\,d]\,b\,b\,a 
       \xrightarrow{\;ad\to abb\;} a\,b\,b\,b\,b\,a
\end{align*}
    By induction on $n\geqslant 1$, it is easy to show that $a\,b^n\,a\ \to^+\ a\,b^{n+1}\,a$, where the sequence
    starts by applying the rule ``$b\,a \to d\,a$" first and once, followed by $(n-1)$ applications of the 
    rule ``$b\,d \to d\,b$", followed by a single application of the rule ``$a\,d \to a\,b\,b$", for a total of $(n+1)$ 
    steps. We thus have an infinite rewrite sequence $\alpha$ of the form:
    \[
    \alpha: \ \ a\,b\,a \ \xrightarrow{\text{2 steps}}\ a\,b\,b\,a
            \ \xrightarrow{\text{3 steps}}\ a\,b\,b\,b\,a 
            \ \xrightarrow{\text{4 steps}}\ \cdots 
            \ a\,b^n\,a\ \xrightarrow{n+1\ \text{steps}}\ a\,b^{n+1}\,a
            \ \cdots 
    \]
    The existence of the infinite sequence $\alpha$ proves that system $T$ is non-terminating. 

    We next show that $\alpha$ does not contain any loop.
    Observe that no rule in $R$ creates or destroys letter $a$. The total count of occurrences of  
    letter $a$ is $2$ in every word $w$ in the sequence $\alpha$, with one occurrence as the \emph{first} and \emph{last}
    letter in $w$, so that every $w$ is of the form $w = a\,w'\,a$ where $w'\in \Set{b,d}^+$. 
    For a loop $w \rightarrow^+ p\,w\,q$ to exist, $p\,w\,q$ must capture both occurrences of $a$ at 
    the two far ends of $w$, which forces the flanking strings $p$ and $q$ to be empty strings. And if
    $p$ and $q$ are not empty strings, then $a$ must also occur as the first letter in $p$ and the last 
    letter in $q$. Hence, no such loop
    can exist in the sequence $\alpha$. System $T$ is non-looping on the infinite rewrite $\alpha$.
    In fact, system $T$ satisfies a stronger result (not shown here): every infinite rewrite sequence
    admitted by $T$ is loop-free.
\end{example}

 
\subsection{Finite Complete String-Rewriting Systems for One-Relator Monoids}
\label{subsect:one-relator}

Before we state open problems related to \emph{one-relator monoids}, we recall definitions with
examples.%
    \footnote{
    $\phantom{s}$ Elaboration of these definitions and examples are in Mark Lawson's book \cite{lawson2026}.
    }

\begin{definition}{Monoids and Relators}
\label{def:monoid}
  A \emph{monoid} can be formally defined via a presentation $\langle \Sigma \mid R \rangle$, 
  where $\Sigma$ is a collection of generators and $R$ is a collection of relations 
  (equations evaluating equivalence between words). The number of \emph{relators} 
  is the number of distinct equations in $R$.
\end{definition}

\begin{examplesansQED}
\label{exa:bicycle}
    Perhaps the most famous and intuitive example of a one-relator monoid is the so-called
    \emph{Bicycle Monoid} (often called the \emph{Bicyclic Semigroup},
    closely related to the so-called \emph{Polycyclic Monoid of rank 1}).
     Its presentation
    can be given as $\langle {a, b} \mid {a\cdot b = 1} \rangle$. There are two significant aspects about this monoid:
    \begin{itemize}
        \item 
        This system utilizes two generators, $a$ and $b$, bound by a single relation stating that 
        multiplying $a$ by $b$ collapses the string into the identity element $1$ (representing
        the empty word $\varepsilon$). 
        \item 
        Crucially, the relation is non-commutative, \ie, $b\cdot a \neq 1$. Because the word $b\cdot a$ cannot be reduced, 
        this one-relator monoid is infinite. Every element in the monoid can be uniquely written in the irreducible
        normal form $b^i\cdot a^j$ for non-negative integers $i, j \geqslant 0$. This structures the monoid to model 
        shifting operations on a one-sided infinite tape.
        \hfill \QED
    \end{itemize}
\end{examplesansQED}

\begin{examplesansQED}
\label{exa:half-grid}
    A structurally clean example of a two-relator monoid is the so-called
    \emph{Monoid of the Half-Grid} (geometrically representing a cylinder grid).
    Its presentation can be given as $\langle {a, b} \mid {a\cdot b = b\cdot a , \ a\cdot a = 1} \rangle$. 
    Significant aspects about this monoid are:
\begin{itemize}
    \item This presentation introduces two independent structural equations:
    \begin{enumerate}
        \item Relation 1, $a\cdot b = b\cdot a$, enforces that the generators commute, meaning the evaluation order 
              of $a$ and $b$ is symmetric.
        \item Relation 2, $a\cdot a = 1$, enforces an involution property on the generator $a$, stating that 
              applying $a$ twice annihilates it back to the identity.
    \end{enumerate}
    \item In this system, the generator $a$ acts like a 180-degree reflection flip 
    (doing it twice brings the system back to its origin), while $b$ acts like a continuous forward 
    translation step. Due to the two constraints, every element in this monoid can be uniquely reduced
    to the normal form $b^n$ or $a\cdot b^n$ for $n \geqslant 0$.
    \hfill \QED
\end{itemize}
\end{examplesansQED}

String-rewriting systems are closely connected to presentations of monoids.  A finite monoid presentation
$\langle \Sigma \mid R\rangle$ can be studied by orienting equations into rewrite rules.
In favorable cases, one obtains a finite complete SRS. As string-rewriting systems,
the monoids in Examples~\ref{exa:bicycle} and~\ref{exa:half-grid} can be specified 
by, respectively:
\[
    \big( \Set{a, b}\ ,\ \Set{a\,b \to \varepsilon ,\ \varepsilon \to a\,b} \big) 
    \quad\text{and}\quad
     \big( \Set{a, b}\ ,\ \Set{a\,b \to b\,a ,\ b\,a \to a\,b, \ a\,a\to \varepsilon, \ \varepsilon \to a\,a} \big)\; .
\]
When we convert the monoid relator $a\cdot b = 1$ 
to the string rewrite rules
$\Set{ a\,b \to \varepsilon ,\ \varepsilon \to a\,b }$, it is convenient to omit 
the multiplication (or binary function symbol) denoted by ``$\cdot$".

Recall from \HItxtG{\color{blue}{Subsection~\ref{subsect:complete}}} that a SRS 
is \emph{complete} if it is \emph{globally terminating} and \emph{confluent}. 
In such a system, every word has a unique normal form.  
Therefore the word problem is decidable by reducing both words to normal form:
\[
u \approx v
\quad\Longleftrightarrow\quad
\operatorname{nf}(u)=\operatorname{nf}(v).
\]
A central open problem is the following, where a one-relator monoid is a monoid with a presentation of the form
$\langle \Sigma \mid u=v \rangle$, where there is only one defining relation.

\medskip

\fbox{\parbox{0.985\linewidth}{
\begin{problem}
    \emph{
     Does every one-relator monoid admit a finite complete SRS?
     }
\end{problem}
}}

\medskip

One of the most famous and long-standing open questions is whether 
the word problem is decidable for the class of all
{one-relator monoids} \cite{gray2019}. 
While the overarching problem remains open, it is settled for significant  
subclasses, such as the subclass of one-relator monoids 
of the form $\langle \Sigma \mid w = 1 \rangle$, which was proved to
have a decidable word problem \cite{adjan1966}. 
The preceding problem is closely related to another famous open problem.

\medskip

\fbox{\parbox{0.985\linewidth}{
\begin{problem}
    \emph{
     Does every one-relator monoid have a decidable word problem?
     }
\end{problem}
}}
 
\medskip

If every one-relator monoid admitted a finite complete SRS, then the word problem for every one-relator monoid would be decidable. Hence, resolving the first of the two preceding problems positively will also resolve
the second positively. However, the converse need not hold, since a decidable word problem does not necessarily imply 
the existence of a finite complete SRS.%

\medskip

\Hide{
{\fontsize{13.2}{15} \selectfont 
\textbf{Appendix to
  \HItxtG{\color{blue}{Subsection~\ref{subsect:one-relator}}} --- 
  a digression into the use of AI systems.}
  \par}
  }

{
\subsection*{\Large\emph{Appendix -- 
  a digression into using AI systems }}  
  \addcontentsline{toc}{subsection}{  \qquad 
  {\emph{Appendix --   
  a digression into using AI systems} } }
}

\medskip

{ \setstretch{1.13}
  \small
  \textbf{Beware: Large Language Models (LLMs) and Other AI Systems May Return Nonsense!}
   In the AI field, this phenomenon is widely known as \emph{hallucination},
   where LLMs confidently invent false information, write fake citations,
   or hallucinate historical events.
   And this happens because answers returned by LLMs are based on
   \emph{probabilistic guessing}. LLMs do not ``know'' facts the way humans know them: they predict
   the next most likely word, or phrase, or paragraph, in a sequence based on
   repeated statistical inspections of their training data.
   While the warning is highly accurate, it is important to understand
   the scale of the issue:
   \begin{itemize}
   \item Answers returned by LLMs are not always nonsense.
     True, they excel at summarizing, based on available
     training data, and at writing summaries clearly and idiomatically. 
     But, the danger lies in trusting them completely where errors may have severe consequences.
   \item 
     There is also an extrinsic trap. The incorrect details returned by LLMs are rarely obvious gibberish;
     instead, they are often mixed into highly logical and sophisticated text, 
     making them subtly and 
     deceptively wrong. This makes errors very hard to spot without external fact-checking.
   \end{itemize}

   The upshot is that an LLM should always be used as an \emph{assistant}
   to humans to boost productivity --- never as an \emph{infallible source} of correctness
   and truth.  A ``verify, don't trust'' approach is always the right mindset, especially when the subject
   at hand requires absolute certainty.%
   \footnote{
   $\phantom{s}$
   A natural question students raise 
   since LLMs became widely visible in 2022-23 is the following:
   \emph{if hallucinations stem from false facts in training data, why do developers not simply 
   use cleaner and more carefully vetted data?} This diagnosis is understandable, but incomplete. 
   Recent research shows that a model trained on an accurate, or mostly accurate, corpus may still
   hallucinate. One reason is statistical: reliably generating a correct answer can be harder 
   than recognizing whether a proposed answer is valid. Consider a true fact that appears only 
   once in the training data, such as the exact page range of an obscure paper. 
   The model has only one example from which to learn it, and one example may not be enough 
   to fix the fact reliably in memory, even when the example itself is accurate. This is a 
   sampling limitation, not a data-quality defect: the lone source is true, but truth 
   alone does not guarantee that the fact will be learned correctly. A separate issue arises 
   from how models are graded during training and evaluation. Many benchmarks reward a lucky 
   confident guess, while a confident wrong answer and an honest ``I don’t know" often receive 
   the same score. This scoring regime pushes models toward guessing rather than abstaining, 
   even in cases where abstention would be the more honest and better response for the user. The recent 
   paper \cite{kalai2025} gives a good argument for this point, and shows why the common 
   explanation blaming hallucinations on dirty training data alone is inadequate.
    }

Like most mathematicians (and most academics in general), I have plenty to say about 
interactions between humans and AI systems. Some are positive, some are negative. 
In this appendix, I limited myself to recounting one particular annoying 
incident that occurred while writing this document. My views 
on LLMs and AI systems in general are elsewhere.%
    \footnote{
    $\phantom{s}$
    See, for example, Section 6 in \cite{kfoury2018} from 2018 when I was
    more hopeful about AI systems, and the first of three
    lectures in \cite{kfoury2026} from 2026 when I became increasingly skeptical of
    AI systems unsurpervised by humans.}
   
In my older set of lecture notes on which I based this document, 
I had included counter-examples due
to Squier to the conjecture that a decidable word problem for a
one-relator monoid $M$ implies a finite complete SRS representing $M$.
Squier defined a family of one-relator monoids that have a decidable
word problem without having finite complete SRSs \cite{squier1987, squier1994}. 
In my attempt to have an update, I asked the LLM {ChatGPT} to provide me with
other counter-examples, which I hoped would be more recent than Squier's. On closer scrutiny
(to my surprise) {ChatGPT}'s answer contained errors, despite its apparent 
sophistication with terminology and definitions. I then asked the LLMs {Gemini} and
{Claude} to correct {ChatGPT}'s answer, but (again to my surprise) they basically confirmed
the latter, since they only proposed minor trivial adjustments. 
I could, of course, have pointed out the errors and asked my question again, 
but I did not, as I wanted to keep a record of the wrong answers.
Here is {ChatGPT}'s specious answer, with a part denoted ``[\,$\ldots$\,]" omitted: 
   \begin{quote}
    The Baumslag-Solitar group 
    $\mathrm{BS}(1,2) = \langle a, t \mid t\cdot a\cdot t^{-1} = a^2 \rangle$ 
    is a \emph{metabelian group} (\ie, it has an abelian normal
    subgroup, which is here the dyadic rationals $\mathbb{Z}[1/2]$)
    and is the \emph{semidirect product} of 
    $\mathbb{Z}[1/2]$ by $\mathbb{Z}$, the latter being 
    the group whose elements are
    $\Set{ (b, k) \mid b \in \mathbb{Z}[1/2], \, k \in \mathbb{Z} }$
    and whose binary operation is defined by
    $(b_1\,, k_1) \cdot (b_2\,, k_2) = (b_1 + 2^{k_1} b_2 \, , \, k_1 + k_2)$
    for all elements
    $(b_1,k_1)$ and $(b_2,k_2)$.
    By virtue of this structure, $\mathrm{BS}(1,2)$ has a decidable word
    problem; however, it admits no finite complete SRS. Intuitively:
    the relation $t\cdot a\cdot t^{-1} = a^2$ forces length doubling under
    conjugation, creating irresolvable critical pair cascades regardless
    of orientation. Hermiller and Meier proved this result independently
    using homological methods, later publishing jointly
    \cite{hermillermeier1995}. [\,$\ldots$\,] 
   The dyadic rationals $\ints [1/2]$ are the rational numbers whose
   denominator is a power of~$2$. They form an abelian group, and thus
   a monoid too, under addition ``$+$".
   The complexity of the decidability of the word problem
   for $\mathrm{BS}(1,2)$ is considered in a more recent paper \cite{weiss2018}.
      \end{quote}

   \textbf{Correction, after laborious forensic (human!) effort.} 
   The passage above, although reportedly confirmed
   by three different LLMs, contains a central error and cannot be
   trusted as stated --- it is itself a good illustration of the warning
   at the top of this appendix. The description of $\mathrm{BS}(1,2)$ as
   \emph{metabelian}, with normal subgroup $\ints[1/2]$ and the \emph{semidirect product}
   law displayed above, is correct, and it is true that $\mathrm{BS}(1,2)$
   has a decidable word problem. But the claim that $\mathrm{BS}(1,2)$
   \emph{admits no finite complete SRS} is false. The string-rewriting system
   $(\Sigma,R)$ whose alphabet is $\Sigma = \Set{a, a^{-1}, t, t^{-1}}$ and 
   whose set of rewrite rules is:
   \begin{alignat*}{4}
   & R =\ &&\SET{\, a^{x}\, a^{-x} \to 1 ,\ \
         t^{x} \,t^{-x} \to 1 ,\ \
         a^2\, t \to t\, a ,\ \
         a^{-1}\, t \to a\, t\, a^{-1} ,\ \
         a^{x}\, t^{-1} \to t^{-1}\, a^{2x}
         \ \Big|\ x \in \Set{+1, -1} \,}    
         \\[.8ex]
    & && \quad \text{where $a^{+1} = a$,
            \ $a^{-(+1)} = a^{-1}$,\ \ $a^{-(-1)} = a$,\ \ $a^{2(+1)} = a\,a$,
            \ $a^{2(-1)} = a^{-1} a^{-1} $ }     
   \end{alignat*}
   is terminating and confluent, hence complete. The set $R$ includes $8$ rewrite rules
   which can be shown --- straightforward exercise left to the reader --- to be equivalent
   to the defining relator of $\mathrm{BS}(1,2)$.
   This system appears in work of Brittenham and Hermiller on
   stackable groups \cite{brittenham2018}, building on the earlier
   rewriting-systems literature for solvable Baumslag--Solitar groups
   surveyed there. In particular, the relation $t\cdot a \cdot t^{-1}=a^2$
   does \emph{not} force ``irresolvable critical pair cascades'':
   the relevant overlap is resolved cleanly by the rules $a^2\,t \to t\,a$
   and $a^{-1}\,t \to a\,t\,a^{-1}$ above, and completion terminates.
   Since $\mathrm{BS}(1,2)$ admits a finite complete SRS, it automatically
   has \emph{finite derivation type} (a property further elaborated in 
   \HItxtG{\color{blue}{Subsection~\ref{subsec:FDT}}} below), by the Squier--Otto--Kobayashi theorem
   \cite{squier1994}. So $\mathrm{BS}(1,2)$ is not a counterexample to
   the conjecture at all --- if anything, it illustrates the opposite
   direction.

   What $\mathrm{BS}(1,2)$ \emph{is} a genuine, well-known counterexample
   for what are called \emph{automaticity} and \emph{almost convexity}, 
   properties distinct from (though related to) admitting a finite complete SRS:
   it is not \emph{automatic}, since $\mathrm{BS}(m,n)$ is \emph{automatic} only when $m=n$
   (see Section 7.4 in \cite{epstein1992}), and it is not \emph{almost convex} \cite{millershapiro1997}.
   It seems likely that the passage above is a conflation of this
   genuine failure (of \emph{automaticity}/\emph{almost convexity}) with the
   unrelated and false claim about finite complete SRSs.

   The attribution to Hermiller and Meier in the quoted passage above
   is also not right. They did
   publish jointly on rewriting systems for groups of exactly this
   kind, but their result is a \emph{positive} one (constructing tame
   combings and rewriting systems for solvable Baumslag--Solitar groups
   and similar families), not a nonexistence proof
   \cite{hermillermeier1997}. From what I could gather, 
   they have not published results that could be misinterpreted as showing
   $\mathrm{BS}(1,2)$ lacks a finite complete SRS, and the direction of
   their actual work is the reverse of what the passage claims. Note
   also that the bibliography entry originally supplied for this
   citation, \cite{hermillermeier1995}, does not match any paper about
   $\mathrm{BS}(1,2)$; the real paper matching those author names and
   that year is \emph{Algorithms and geometry for graph products of
   groups}, Journal of Algebra 171(1):230--257, 1995, which is about
   graph products, not Baumslag--Solitar groups --- itself a symptom
   of the same fabricated-citation phenomenon the warning above cautions
   against.

   For a genuine, citable example of a one-relator (indeed finitely
   presented) \emph{group} with decidable word problem and provably no
   finite complete SRS, see Stallings' group \cite{stallings1963}, which
   fails the homological finiteness property $FP_3$ and is the
   group-theoretic analogue of Squier's monoid examples.
   \par
   }


\subsection{Characterizing Monoids with Finite Complete Presentations}

A finite complete SRS is one of the most useful restricted forms of a string-rewriting
system.  It gives canonical normal forms and an effective solution to the word problem. A broad
research direction is threfore the following.

\medskip

\fbox{\parbox{0.985\linewidth}{
\begin{problem}
    \emph{Characterize the finitely presented monoids that admit finite complete string-rewriting systems.
    }
\end{problem}    
}}

\medskip

This is not merely a technical problem about rewriting algorithms.  It asks for an intrinsic algebraic 
characterization of those monoids for which finite canonical rewriting is possible. Some necessary conditions are known.  
For example, the Anick-Groves-Squier theorem introduced homological and homotopical finiteness conditions 
associated with finite complete SRSs.%
      \footnote{
      $\phantom{s}$ Further details regarding this paragraph are in many publications by Kenneth S. Brown. 
      See, in particular, \cite{brown1992} and the very recent \cite{gray2026}.
      }
In particular, finite complete SRSs imply certain finiteness properties, 
such as \emph{Finite Derivation Type} (elaborated in the next section).  
However, these conditions do not yet provide a complete characterization.


\subsection{Finite Derivation Type (FDT)}
\label{subsec:FDT}

The property of having a \emph{Finite Derivation Type} (FDT) is a  
homological property for monoids and string rewriting systems. 
It was introduced by Craig Squier to capture a geometric form of ``finiteness'' 
that goes beyond having a decidable word problem or a finite presentation \cite{squier1994}.

As indicated already, if a monoid can be presented by a finite complete SRS, 
then its word problem is decidable. However, it turns out that some finitely presented
monoids possess a decidable word problem but \emph{cannot} be presented by any finite complete 
SRS. To explain this restriction, Squier analyzed the graphs of paths (or derivations) between equivalent words.
FDT formalizes the condition that the ``loops'' and ``holes'' formed by distinct rewriting pathways 
between identical words can be completely bounded and filled in (\ie, homotopically trivialized) by a finite 
set of foundational relations. The preceding comments motivate the following definitions.

\begin{definition}{Derivations and Paths}
\label{def:derivations}
   Let $T=(\Sigma, R)$ be a finite SRS. 
   The system $T$ defines a monoid $M = \Sigma^* / \leftrightarrow_R^*$. 
   We can model the structure of $M$ by a \emph{rewriting graph}  
   denoted $\Gamma(\Sigma, R)$:
\begin{itemize}
    \item $V$ is its set of \emph{vertices}, consisting of all the words in the free monoid $\Sigma^*$,
    \item $E$ is its set of \emph{edges}, consisting of all the single-rewrite steps $\ell \rightarrow r$ and their 
     inverses $r \leftarrow \ell$. An edge $e$ connecting $\ell$ and $r$ is written as 
    $e: \ell \leftrightarrow r$.
\end{itemize}
A \emph{derivation} (or \emph{path}) $\alpha$ is a finite sequence of adjacent edges connecting 
a starting word $u$ to an ending word $v$. If two distinct paths $\alpha$ and $\beta$ share the same 
source word $u$ and target word $v$, they form a \emph{parallel path pair}, 
which can be closed into a structural loop.
\end{definition}

\begin{definitionsansQED}{Homotopy Relation on Paths}
\label{def:homotopy}
    Let $T=(\Sigma, R)$ be a finite SRS. 
    We define an equivalence relation $\sim$ (called a \emph{homotopy}) on the paths of its
    rewriting graph $\Gamma(\Sigma, R)$. Two paths are equivalent if they differ only by:
\begin{enumerate}
    \item \emph{Trivial loops:} Traversing an edge $e$ and immediately backtracking along its 
    inverse $e^{-1}$, \ie,  $e \cdot e^{-1} \sim \varepsilon\ \text{(empty path)}$.
    \item \emph{Orthogonal independence:} If two non-overlapping substrings are 
    rewritten independently within a single word (\eg, rewriting $a \rightarrow b$ 
    at the head of a string and $c \rightarrow d$ at the tail), the execution order 
    is symmetric. The paths form a diamond shape, and the two tracks around this diamond 
    are declared \emph{homotopically equivalent}.
    \hfill \QED
\end{enumerate}
\end{definitionsansQED}

\begin{definition}{The FDT Property}
\label{def:FDT}
    Let $T=(\Sigma, R)$ be a finite SRS and $\Gamma(\Sigma, R)$ its induced
    rewriting graph. Let $C$ be a set of closed loops (or parallel path pairs) in $\Gamma(\Sigma, R)$. 
    We extend our homotopy relation $\sim$ to another equivalence relation $\sim_C$ 
     by contracting every loop in $C$ to a point (\ie, traversing one 
     half of the loop is equivalent to traversing the other half).

    The system $T = (\Sigma, R)$ is defined to have \emph{Finite Derivation Type} (FDT) 
    if there exists a \emph{finite set $C$} of path pairs such that for \emph{every} parallel pair 
    of paths $\alpha, \beta$ in $\Gamma (\Sigma,R)$, we have:
    \[ \alpha \sim_C \beta \]
    Geometrically, $T$ has FDT if the fundamental group-like structure of its rewriting graph 
    can be made homotopically trivial (simply connected) by ``tiling over'' all structural 
    holes using only a finite set of tiles $C$ \cite{squier1994}.
\end{definition}    

\begin{examplesansQED}   %
This is a very simple example to illustrate Definitions~\ref{def:derivations}, \ref{def:homotopy},
and~\ref{def:FDT}: the cyclic monoid of order 2, which is in fact the cyclic group of order 2, 
namely, $\mathbb{Z}_2$.%
     \footnote{
     $\phantom{s}$ If you need to refresh your basic knowledge of \emph{cyclic groups}, the 
     {\href{https://en.wikipedia.org/wiki/Cyclic_group}{\color{blue}{Wikipedia page}}}
     will give you enough (and more) for this handout.
     }

Let $T = (\Sigma, R)$ be a finite string-rewriting system where
\( \Sigma = \Set{a} \) and $R = \Set{ a\,a \rightarrow \varepsilon }$. 
The two-element monoid presented by $T$ is $M = \{\varepsilon, a\}$.

\begin{itemize}
    \item 
    \emph{Derivations and Paths:}
    The rewriting graph $\Gamma(\Sigma, R)$ contains an infinite set of vertices 
    $V = \{\varepsilon,\ a,\ a\,a,\ a\,a\,a,\ \dots\}$ corresponding to all the elements of the free monoid $\Sigma^*$. 
    Consider a source word $u = a\,a\,a\,a$ and a target word $v = \varepsilon$. 
    We can define two distinct parallel paths (or derivations) $\alpha$ and $\beta$ connecting $u$ to $v$:
\begin{align*}
    \alpha &:\ [a\,a]\,a\,a \ \xrightarrow{aa\,\rightarrow\,\varepsilon}\ \varepsilon\, [a\,a] 
    \ \xrightarrow{aa\,\rightarrow\,\varepsilon}\ [\varepsilon\,\varepsilon]\ =\ \varepsilon
    \\[1ex]
    \beta &:\ a\,a\,[a\,a]\ \xrightarrow{aa\,\rightarrow\,\varepsilon} \ [a\,a] \,\varepsilon
    \ \xrightarrow{aa\,\rightarrow\,\varepsilon}\ [\varepsilon\,\varepsilon]\ =\ \varepsilon
\end{align*}
Because $\alpha$ and $\beta$ share the same source $a\,a\,a\,a$ and target 
$\varepsilon$, they form a \emph{parallel path pair}, closing a structural loop in $\Gamma(\Sigma, R)$.
    
    \item \emph{Homotopy Relation:}
    The parallel paths $\alpha$ and $\beta$ exemplify \emph{orthogonal independence}.
    In the word $a\,a\,a\,a$, the first two $a$'s and the last two $a$'s are non-overlapping substrings: 
    \begin{itemize}
      \item path $\alpha$ executes the rewrite at the head of the string first, then at the tail.
      \item path $\beta$ executes the rewrite at the tail of the string first, then at the head.
    \end{itemize}
    Because the execution order of these independent steps is symmetric, the paths track around a 
    diamond shape in the graph geometry. By Definition~\ref{def:homotopy}, they are declared %
    homotopically equivalent ($\alpha \sim \beta$).

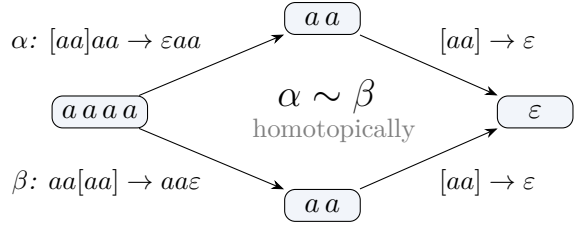
\begin{figure}[h]
    \qquad\qquad
    \begin{tikzpicture}[
        node distance=0.8cm and 1.8cm,   
        vertex/.style={draw, rectangle, rounded corners, minimum width=1cm, minimum height=0.3cm, fill=blue!5, font=\ttfamily},
        edge label/.style={font=\small\itshape, color=black}
    ]
        \node[vertex] (source) {$a\,a\,a\,a$};
        \node[vertex] (top) [above right=of source] {$a\,a$};
        \node[vertex] (bottom) [below right=of source] {$a\,a$};
        \node[vertex] (target) [below right=of top] {$\varepsilon$};

        \draw[{Stealth}-] (top) -- (source) 
            node[midway, above left, edge label] {$\alpha$: $[aa]aa \rightarrow \varepsilon aa$};
        \draw[-{Stealth}] (top) -- (target) 
            node[midway, above right, edge label] {$[aa] \rightarrow \varepsilon$};

        \draw[{Stealth}-] (bottom) -- (source) 
            node[midway, below left, edge label] {$\beta$: $aa[aa] \rightarrow aa \varepsilon$};
        \draw[-{Stealth}] (bottom) -- (target) 
            node[midway, below right, edge label] {$[aa] \rightarrow \varepsilon$};

        \node at (3.0,0.2) {\Large $\alpha \sim \beta$};
        \node[font=\small, gray] at (3.1,-0.25) {homotopically};

    \end{tikzpicture}
\end{figure}


    \item \emph{The FDT Property:}
    While orthogonal independence handles non-overlapping rewrites, the graph $\Gamma(\Sigma, R)$ 
    also possesses structural loops caused by overlapping rules. Consider the word $a\,a\,a$. 
    We can find two competing, overlapping applications of the rule $a\,a \rightarrow \varepsilon$:
    \[
    \gamma_1 : [a\,a]\,a \rightarrow a \quad\text{and}\quad
    \gamma_2 : a\,[a\,a] \rightarrow a\ .
    \]
   This generates a parallel path pair (and thus a loop) starting at $a\,a\,a$ and ending at $a$. 

   To demonstrate that $T$ has FDT, we define a finite set $C$ 
   containing exactly this critical overlapping loop instance. By adjoining this single loop 
   tile to our homotopy relation, any other parallel paths generated by deeper string 
   extensions can be successfully contracted to a point via $\sim_C$. Since a finite set 
   $C$ is sufficient to make the entire group-like architecture of $\Gamma(\Sigma, R)$ 
   homotopically trivial, the system $T$ possesses FDT. 
   \hfill \QED
\end{itemize}
\end{examplesansQED}
  
\paragraph{Squier's Theorem:}
\emph{If a monoid $M$ can be presented by a finite complete string-rewriting system, 
then $M$ must have a Finite Derivation Type} \cite{squier1994}.

Because FDT is an invariant, if a monoid can be shown to lack FDT, it is mathematically impossible 
to ever construct a finite complete SRS for it. 
This insight provides an absolute boundary for automated completion workflows, most notably the 
\emph{Knuth-Bendix completion algorithm} \cite{baader+nipkow, knuthbendix1970}. When an equational theory 
fails the FDT condition, the completion procedure is guaranteed to diverge or fail to produce 
a finite canonical rewrite set.
A finite complete SRS implies FDT, but the converse implication does not hold. 
This leads to the following question.

\medskip

\fbox{\parbox{0.985\linewidth}{
\begin{problem}
    \emph{To what extent does Finite Derivation Type characterize monoids admitting finite complete 
    string-rewriting systems? Is there any homotopy or homological finiteness condition, stronger than FDT,
    which characterizes monoids admitting finite complete string-rewriting systems?
    }
\end{problem}    
}}

\medskip

\subsection{Closure Properties of Finite Complete String-Rewriting Systems}

Another family of open problems concerns closure properties.
For example, suppose a monoid or group has a finite complete string-rewriting system (SRS). 
Does a naturally associated substructure also have one?  Conversely, 
if a finite-index substructure has a finite complete SRS, does the larger structure have one?%
     \footnote{
     $\phantom{s}$ Basic definitions related to the \emph{index of a substructure} can be
     found at the 
     {\href{https://en.wikipedia.org/wiki/Index_of_a_subgroup}{\color{blue}{Wikipedia page}}}.
     }

Questions of this kind ask whether finite complete SRSs 
are preserved under standard algebraic constructions. A representative example is the following.

\medskip

\fbox{\parbox{0.985\linewidth}{
\begin{problem}
    \emph{
    Is the property of admitting a finite complete SRS preserved under 
    taking certain finite-index subgroups or finite-index extensions?
    }
\end{problem}
}}

\medskip

Many special cases are known, but the general closure landscape remains largely open. Relevant 
references are \cite{cairnes2000, grovessmith1993, pridewang2000}.

\subsection{Beyond the Word Problem: Conjugacy, Cyclic Equality, Green's Relations}

While a finite complete SRS guarantees an effective solution to the standard word problem
by mapping every element to its unique irreducible normal form, it does not automatically
render every decision problem over a monoid decidable. Decisions concerning more complex
structural or existential relations --- such as \emph{conjugacy-like} problems, 
\emph{cyclic equality}, or \emph{ideal structure relations} --- frequently remain undecidable. The core computational barrier is that
these relations involve existential quantifiers ($\exists$) over the entire, often infinite, free monoid, converting a deterministic word-reduction task into a complex, unbounded search problem
\cite{book2012string}.

In group theory, two elements $x$ and $y$ are conjugate if there exists an element $g$ such that 
$g^{-1}\cdot x\cdot g = y$. Because a monoid lacks structural inverses, this single relation splits
into several non-equivalent definitions. Two common and widely investigated generalizations
are \emph{primary conjugacy}%
    \footnote{
     $\phantom{s}$
    {Primary conjugacy} in monoids is usually credited to Lyndon and Sch\"utzenberger \cite{lyndon1962}.
    }
and \emph{Otto's conjugacy}. The latter turned out to be equivalent to
\emph{Squier's conjugacy}.%
    \footnote{
    $\phantom{s}$
    Friedrich Otto (1984)
    formulated his definition in the context of formal language theory \cite{otto1984}. 
    He used it to investigate the complexity of the conjugacy problem within rewriting 
    structures --- specifically, string-rewriting systems that satisfy the confluence property. Craig Squier (1987) approached the exact same equation from the perspective of homological algebra and semigroup theory \cite{squier1987}. Squier utilized his definition while mapping out how word rewriting interacts with chains of module resolutions, establishing foundational 
    boundaries for homological finiteness conditions.
    }
    
\begin{definition}{Primary Conjugacy}
\label{def:primary}
Let $M$ be a monoid. Two elements $x, y \in M$ are said to be \emph{primary conjugate} 
(here denoted $x \sim_p y$) if there exist elements $u, v \in M$ such that:
$x = u\cdot v$ and $y = v\cdot u$.
\end{definition}

\emph{Primary conjugacy} 
is reflexive and symmetric, but it is \emph{not} generally transitive in an arbitrary monoid. To form a proper equivalence relation, we evaluate its transitive closure, $\sim_p^*$.

\begin{definition}{Otto-Squier Conjugacy}
\label{def:Otto-Squier}
Let $M$ be a monoid. Two elements $x, y \in M$ are said to be conjugate in the sense of Otto-Squier (here denoted $x\ {\sim}_{o,s}\ y$) if there exist elements $g, h \in M$ such that: $x\cdot g = g\cdot y$ and $y\cdot h = h\cdot x$.
\end{definition}

Again here, \emph{Otto-Squier conjugacy} is not an equivalence relation on an arbitrary monoid
because it can fail transitivity. An equivalence relation is obtained by taking its transitive
closure, ${\sim}_{o,s}^*$.

\begin{fact}
    In Definitions~\ref{def:primary} and~\ref{def:Otto-Squier}, if $M$ is a group (not only a monoid), then 
    both \emph{primary conjugacy} and \emph{Otto-Squier conjugacy} collapse into standard group conjugacy.
\end{fact}

\begin{remark}
In Definition~\ref{def:Otto-Squier} of \emph{Otto-Squier conjugacy},
if the monoid $M$ contains an absorbing zero element ($0$), it is conventionally required that the 
connecting elements $g$ and $h$ be non-zero (\ie, $g\neq 0$ and $h \neq 0$) to avoid trivial conjugacy classes.
In the presence of a zero element, if we allow $g = 0$ and $h = 0$, then
we can plug them into the equations for any arbitrary pair of elements $x$ and $y$, \ie, 
every single element in the monoid becomes conjugate to every other element, and the entire
monoid becomes a single trivial conjugacy class. Further details are in \cite{araujo2011, otto1984}.
\end{remark}

When SRSs are finite and complete, the conjugacy problem remains undecidable
in general, \ie, there is no algorithm which, when given an arbitrary finite and
complete SRS, can decide whether the bridging elements $u,v$ 
(in Definition~\ref{def:primary}) or $g,h$ (in Definition~\ref{def:Otto-Squier}) exist \cite{araujo2017}.

When monoids are restricted to the subclass of
groups, we have the following facts. Although the conjugacy problem for all one-relator groups 
is a long-standing open problem --- it is not known whether it is decidable or not --- 
the problem is decidable and completely solved for several major subclasses of 
one-relator groups, such as the subclass of one-relator groups with torsion \cite{magnus1966}.%
    \footnote{$\phantom{s}$
    Further information about one-relator groups can be found at the long
    {\href{https://en.wikipedia.org/wiki/One-relator_group}{\color{blue}{Wikipedia page}}}, which includes
    many of the classic references.
    }
However, this does not automatically translate to monoids.

\medskip

\fbox{\parbox{0.985\linewidth}{
\begin{problem}
    \emph{
   Let $M = \langle \Sigma \mid w = 1 \rangle$ be a special one-relator monoid.
   \begin{enumerate}
    \item Is the primary conjugacy problem (deciding if $x \sim_p^* y$) decidable for all such monoids?
    \item Is the Otto--Squier conjugacy problem (deciding if $x \sim_{o,s} y$) decidable for all such monoids?
   \end{enumerate}
    }
\end{problem}
}}    
 
\medskip  

It is well established that termination and confluence together (completeness) are sufficient (but not 
necessary) for the word problem for monoids to be decidable.
For conjugacy problems, however, the independence of these properties is less understood. 
Ara\'ujo et al.\ proved that undecidability can persist under a \emph{finite complete} SRS \cite{araujo2017}. 
Conversely, if we restrict SRSs to specific geometric or algebraic families, we get positive results. 
For example, if a monoid can be presented by a finite \emph{special} complete SRS (where all the rewrite 
rules are of the form $\ell \rightarrow 1$), the conjugacy problems collapse and become decidable via cyclic 
string shifts \cite{otto1984}. This leaves a huge gap between arbitrary complete SRSs and special complete SRSs.

\medskip

\fbox{\parbox{0.985\linewidth}{
\begin{problem}
    \emph{
    Identify the exact syntactic boundaries of a finite complete SRS that guarantee the decidability of primary conjugacy 
    and/or Otto-Squier conjugacy. Specifically:
\begin{enumerate}
    \item Is conjugacy (primary or Otto-Squier) decidable for monoids presenting a finite complete SRS that is \emph{monadic} (rules $\ell \rightarrow r$ where $|\ell| > |r|$ and $r \in \Sigma \cup \{1\}$)?
    \item Is conjugacy (primary or Otto-Squier) decidable for monoids presenting a finite complete SRS that is strictly \emph{length-reducing} ($|\ell| > |r|$ for all rules)?
\end{enumerate}
    }
\end{problem}
}}

\medskip
 
Cyclic equality formalizes the algebraic action of taking a word, splitting it, and cyclically swapping the prefixes and suffixes inside a quotient structure.

\begin{definition}{Cyclic Equality}
Let $M$ be a monoid presented by a string-rewriting system $T = (\Sigma, R)$. Two elements $x, y \in M$ 
are defined to be \emph{cyclically equal} if there exists a finite chain of elements 
$z_0, z_1, \dots, z_k \in M$ such that $z_0 = x$, $z_k = y$, and for each index $i$, there exist 
components $u_i, v_i \in M$ such that:
$z_i = u_i v_i$ and $z_{i+1} = v_i u_i$.
\end{definition}

In a free monoid $\Sigma^*$, cyclic equality matches 
cyclic string permutations 
(\eg, $a\,b\,c \sim b\,c\,a \sim c\,a\,b$). However, within a presented monoid $M = \Sigma^*/\leftrightarrow_R^*$, the permutations $u\,v \leftrightarrow v\,u$ heavily interact with the underlying rewrite rules 
$\ell \rightarrow r$. A complete SRS manages the directional $\ell \rightarrow r$ reductions perfectly, but it cannot structurally accommodate the commutative-like cyclic shifting behavior, leaving cyclic equality undecidable for various convergent systems \cite{choffrut1997}.

\medskip

\fbox{\parbox{0.985\linewidth}{
\begin{problem}
    \emph{
     Is the cyclic equality problem decidable for all monoids that can be presented by a finite, complete, 
     and \emph{length-reducing} (or \emph{monadic}) string-rewriting system?
    }
\end{problem}
}
}

\medskip

Beyond conjugacy and cyclic equality, a monoid's internal structure is framed by equivalence relations 
that dictate its ideal structures, known collectively as Green's Relations \cite{green1951}.

\begin{definitionsansQED}{Green's $\mathcal{R}$ and $\mathcal{L}$ Relations}
Let $M$ be a monoid. 
\begin{itemize}
    \item Two elements $x, y \in M$ are $\mathcal{R}$-equivalent ($x\, \mathcal{R}\, y$) if they generate identical principal right ideals, meaning $x\,M = y\,M$:
    $x\, \mathcal{R}\, y \iff \exists u, v \in M \text{ such that } x\,u = y \text{ and } y\,v = x$ .
    \item Two elements $x, y \in M$ are $\mathcal{L}$-equivalent ($x \,\mathcal{L}\, y$) 
    if they generate identical principal left ideals, meaning $M\,x = M\,y$:
    $x\, \mathcal{L}\, y \iff \exists u, v \in M \text{ such that } u\,x = y \text{ and } v\,y = x $ .
    \hfill \QED
\end{itemize}
\end{definitionsansQED}

Just like conjugacy and cyclic equality, deciding Green's relations under a finite complete SRS requires 
evaluating existential conditions across the infinite monoid. While the word problem maps any reduction path to a single terminal node, these structural properties track across complex structural dimensions that an SRS alone cannot compute.

\medskip

\fbox{\parbox{0.985\linewidth}{
\begin{problem}
    \emph{
     Is it decidable whether two elements are $\mathcal{R}$-equivalent (or $\mathcal{L}$-equivalent) in an 
     arbitrary monoid presented by a finite, complete, and \emph{length-reducing} string-rewriting system?
    }
\end{problem}  
}}

\medskip

\subsection{Formal Languages that Are Church-Rosser Congruential}

A string rewrite system $T = (\Sigma , R)$ partitions $\Sigma^*$ into \emph{congruence classes}.
Two strings $u, v\in {\Sigma}^*$ are \emph{congruent} iff $u {\leftrightarrow}^* v$. A subset of 
words $A \subseteq {\Sigma}^*$ is a \emph{congruence class} iff it is a complete equivalence class under 
this relation; that is, for any $u \in A$, $A = \{v \in \Sigma^* \mid u \leftrightarrow^* v\}$.

A string-rewriting system is Church-Rosser if it is confluent (\emph{cf.} 
\HItxtG{\color{blue}{Subsection~\ref{subsect:complete}}}). 
A formal language is called \emph{Church-Rosser congruential} if it can be described as a 
union of finitely many congruence classes of a finite, confluent, length-reducing SRS $T = (\Sigma , R)$.
Note that by being confluent and length-reducing, it is also complete 
(\emph{cf.} \HItxtG{\color{blue}{Subsections~\ref{subsection:length} and~\ref{subsect:complete}}}).

There was a long-standing conjecture that every \emph{regular} language 
is \emph{Church-Rosser congruential}, which 
was settled affirmatively \cite{diekert2015}.
A primary benefit of this property is that the language's parsing problem can be 
solved in deterministic linear time and linear space through pure string reduction 
to unique normal forms \cite{mcnaughton1988}.
Furthermore, it translates complex automata-theoretic configurations into purely algebraic properties, 
facilitating direct proofs for structural decision problems like language emptiness and 
equivalence \cite{diekert2015}.

\medskip

\fbox{\parbox{0.985\linewidth}{
\begin{problem}
    \emph{  
    This problem consists of two interdependent parts:
\begin{itemize}
    \item Is there a natural, structurally defined subclass of the deterministic context-free languages (DCFL) that strictly extends the class of all regular languages, such that all members of this subclass are Church-Rosser congruential?
    \item Is there an algorithm to decide whether an arbitrary DCFL belongs to this subclass --- or, equivalently, is it decidable whether an arbitrary DCFL can be presented by a finite, confluent, and length-reducing string-rewriting system with finitely many congruence classes?
\end{itemize}
    }
\end{problem}  
}
}

\subsection{The Madlener and Otto Conjecture}

In abstract algebra, there is a large corpus of work based on using
string-rewriting systems to present groups. An open question in this area of
mathematics concerns \emph{plain groups}, \ie,
groups that are isomorphic to a free product
of finitely many finite groups and finitely many copies of the infinite cyclic group.
It is known that every plain group can be presented by a finite, complete
(confluent + terminating), {length-reducing} string-rewriting system \cite{dietrich2022}. However, the
converse implication --- sometimes called the \emph{Madlener and Otto conjecture} \cite{madlener1987}) ---
remains an elusive open problem.

\medskip

\fbox{\parbox{0.985\linewidth}{
\begin{problem}
    \emph{ 
    Is it the case that if a group $G$ can be presented by a finite, complete
    (confluent + terminating), length-reducing string-rewriting system, then $G$ is a plain group.
    }
\end{problem}
}}

\medskip

 Recent progress by Elder and Piggott \cite{elder2023piggott} established a geometric framework for such 
 SRSs, but the full conjecture remains open. If proven true, it would be
 an algebraic characterization  (``\emph{free products of finitely many finite 
 groups and finitely many copies of the infinite cyclic group}") of all the groups
 that can be presented by a finite, complete, length-reducing SRS.
 
\subsection{Confluence of Length-Reducing Rules with Commutation}

The boundaries of confluence become significantly more complex when standard string-rewriting rules 
are forced to interact with equational axioms, such as commutativity. A landmark result by Narendran and Otto \cite{book2012string, narendran1988} established that it is undecidable whether an arbitrary 
string-rewriting system is \emph{confluent}%
    \footnote{
    $\phantom{s}$
    Articles from the 1980s and 1990s sometimes called confluent SRSs \emph{preperfect}.
    }
if it consists of a 
finite set of strictly length-reducing rules supplemented by even a single commutation rule 
($a\,b \leftrightarrow b\,a$). This undecidability stems from the fact that a length-preserving 
commutation rule introduces non-terminating cyclic shuffles that can bypass 
the combinatorial constraints of length-reduction. 

However, this classic result leaves a structural question regarding system perturbation. 
For example, if we narrow our scope from an arbitrary system to one that is already confluent, 
the behavioral threshold shifts. Specifically, if we start with a finite, strictly length-reducing
system that is explicitly given to be confluent, the operational consequences of 
injecting a single commutative rule remain unknown. This motivates open problems of the following
kind.

\medskip

\fbox{\parbox{0.985\linewidth}{
\begin{problem}
    \emph{ 
    Let $R$ be a finite, strictly length-reducing SRS over $\Sigma$ that is 
    confluent. Let $C = \{a\,b \leftrightarrow b\,a\}$ be a commutation rule 
    for distinct letters $a, b \in \Sigma$, and let $R' = R \cup C$.
\begin{enumerate}
    \item Is it decidable whether the perturbed system $R'$ remains confluent modulo the equational theory induced by $C$?
    \item Characterize precise syntactic conditions on the rules of $R$ under which the introduction of the 
    single commutation rule $a\,b \leftrightarrow b\,a$ preserves the uniqueness of normal forms modulo $C$.
\end{enumerate}
}
\end{problem}
}}

\medskip

\subsection{Optimal Derivation Bounds for Length-Preserving Systems}

The following question was raised in the 1980s \cite{metivier1985, rta20}: Given a single length-preserving rewrite 
rule (such as $b\,a \rightarrow a\,b$), what is the maximum number of steps a rewrite sequence can take before 
looping or stopping? Is it $\mathcal{O}(n^2)$, or is it $\mathcal{O}(n^k)$, or is it some other 
bound based on $n$ and $k$, where $n$ is the size of the initial word in the sequence and $k$ 
is the size of the rewrite rule?

The answer was determined a decade later \cite{bertrand1994}: the upper bound is 
$n^2 / 4$ and does not depend on $k$. And this bound is tight, \ie, it is reached,
when the string-rewriting system $(\Set{a,b},\ \Set{b\,a\to a\,b})$ is used
to reduce the word $b^{n/2} a^{n/2}$. With the single-rule problem now solved,
attention has shifted to generalizations, such as the following.

\medskip

\fbox{\parbox{0.985\linewidth}{
\begin{problem}
    \emph{ 
    Find tight, generalized polynomial bounds for the derivation length of subclasses of
    finite \textbf{multi-rule} (with two or more rewrite rules) length-preserving systems.
    }
\end{problem}
}}


\newpage 

\begin{thebibliography}{9}

{
\setstretch{1.13}
\small

\bibitem{adjan1966}
S.~I.~Adjan. 
\emph{Defining relations and algorithmic problems for groups and semigroups.} 
Proceedings of the Steklov Institute of Mathematics, Number 85, 1966.

\bibitem{araujo2011}
J.~Ara\'ujo, M.~Kinyon, and J.~Konieczny. 
\emph{Conjugacy in Semigroups.} 
Journal of Algebra and Its Applications, Volume 10, Issue 5, pp. 917--933, 2011.

\bibitem{araujo2017}
J.~Ara\'ujo, W.~Bentz, J.~Konieczny, and A.~Malheiro. 
\emph{The conjugacy problem for monoids defined by finite complete presentations.} 
arXiv preprint, arXiv:1703.00027, 2017.

\bibitem{baader+nipkow}
Franz Baader and Tobias Nipkow.
\emph{Term Rewriting and All That}, Cambridge University Press, 1998.

\bibitem{bauer1984}
Gerhard Bauer and Friedrich Otto.
\emph{Finite Complete Rewriting Systems and the Complexity of the Word Problem,}
{Acta Informatica}, 1984.

\bibitem{bertrand1994}
A. Bertrand.
\emph{Sur une conjecture d'Yves Métivier.}
Theoretical Computer Science, 123(1):21--30, 1994.

\bibitem{book2012string}
R. V. Book and F. Otto.
\emph{String-rewriting systems.}
Springer Science \& Business Media, 2012.

\bibitem{brittenham2018}
Mark Brittenham and Susan Hermiller.
\emph{HNN extensions and stackable groups.}
Groups, Geometry, and Dynamics, Volume 12, Issue 3, pp. 1123--1158, 2018.

\bibitem{brown1992}
Kenneth S. Brown.
\emph{The geometry of rewriting systems: a proof of the Anick-Groves-Squier theorem}. in Algorithms and
      Classiﬁcation in Combinatorial Group Theory (Berkeley, CA, 1989), volume 23 of Math. Sci. Res. Inst. Publ. 
      (Springer, New York, 1992), 137–163.

\bibitem{cain2013}
Alan Cain and Victor Maltcev.
\emph{Monoids \(\mathrm{Mon}\langle a,b : a^\alpha b^\beta a^\gamma b^\delta a^\varepsilon b^\varphi=b\rangle\) admit finite complete rewriting systems,} 2013.

\bibitem{cairnes2000}
C. Cairnes, N.~D. Gilbert, and N. Ru\v{s}kuc. 
\emph{Finite derivation type for semigroups with a finite Rees index subsemigroup.} 
Journal of Algebra, Volume 225, Issue 2, pp. 797--812, 2000. 
DOI: \texttt{10.1006/jabr.1999.8147}

\bibitem{caron1991}
A. Caron. 
\emph{Linear Bounded Automata and String Rewriting Systems.} 
Information Processing Letters, Volume 38, Issue 3, pp. 115--119, 1991. 
DOI: \texttt{10.1016/0020-0190(91)90224-R}

\bibitem{choffrut1997}
C.~Choffrut and J.~Karhum\"aki. 
\emph{Combinatorics of Words.} 
Handbook of Formal Languages, Vol. 1: Word, Language, Grammar, Springer, pp. 329--438, 1997.

\bibitem{elder2023piggott}
M. Elder and A. Piggott.
\emph{On groups presented by inverse-closed finite confluent
              length-reducing rewriting systems.}
Journal of Algebra, Vol. 627, pp. 106--131, 2023. 
DOI: \texttt{10.1016/j.jalgebra.2023.03.022}              
%
\Hide{ 
 @article {MR4574102,
    AUTHOR = {Elder, Murray and Piggott, Adam},
     TITLE = {On groups presented by inverse-closed finite confluent
              length-reducing rewriting systems},
   JOURNAL = {J. Algebra},
  FJOURNAL = {Journal of Algebra},
    VOLUME = {627},
      YEAR = {2023},
     PAGES = {106--131},
      ISSN = {0021-8693,1090-266X},
   MRCLASS = {20E06 (20F65 68Q42)},
  MRNUMBER = {4574102},
MRREVIEWER = {Peng\ Choon\ Wong},
       DOI = {10.1016/j.jalgebra.2023.03.022},
       URL = {https://doi.org/10.1016/j.jalgebra.2023.03.022},
}


}

\bibitem{dershowitz2005open}
N. Dershowitz.
\emph{Open. Closed. Open.}
In International Conference on Rewriting Techniques and Applications, Springer, pp. 1-17, 2005.

\bibitem{diekert2015}
Volker Diekert, Manfred Kufleitner, Klaus Reinhardt, and Tobias Walter.
\emph{Regular Languages are Church-Rosser Congruential.} 
Journal of the ACM, Vol. 62, no. 5, pp 1-20,  
\url{https://doi.org/10.1145/2808227},
doi = {10.1145/2808227}, 2015.

\bibitem{dietrich2022}
Heiko Dietrich, Murray Elder, Adam Piggott, Youming Qiao, and Armin Weiß.
\emph{The isomorphism problem for plain groups is in $\Sigma_3^{\mathsf{P}}$}.
arXiv preprint arXiv:2110.00900, \url{https://arxiv.org/abs/2110.00900}, 2022.

\bibitem{epstein1992}
David B.~A. Epstein, James W. Cannon, Derek F. Holt, Silvio V.~F. Levy,
Michael S. Paterson, and William P. Thurston.
\emph{Word Processing in Groups.}
Jones and Bartlett Publishers, Boston, MA, 1992.

\bibitem{geser2001}
Alfons Geser.
\emph{On the Termination Problem for One-Rule Semi-Thue Systems.}
{Rewriting Techniques and Applications}, 2001.

\bibitem{geser2002}
Alfons Geser.
\emph{Decidability of Termination of Grid String Rewriting Rules.}
SIAM J. on Computing, 2002.

\bibitem{geser-hofbauer-waldmann2002}
Alfons Geser, Dieter Hofbauer, and Johannes Waldmann.
\emph{Loops of Superexponential Lengths in One-Rule String Rewriting.}
{Rewriting Techniques and Applications}, 2002.

\bibitem{geser-hofbauer-waldmann2004}
A. Geser, D. Hofbauer, and J. Waldmann. 
\emph{The Termination Problem for One-Rule Leftmost String Rewriting is Decidable.} 
Journal of Symbolic Computation, Volume 38, Issue 5, pp. 1387--1411, 2004. 
DOI: \texttt{10.1016/j.jsc.2004.06.002}

\bibitem{geser-zentama}
Alfons Geser and Hans Zantema.
\emph{Non-Looping String Rewriting.}
{RAIRO Theoretical Informatics and Applications}, 1999.

\bibitem{gray2019}
R.~D.~Gray. 
\emph{Undecidability of the word problem for one-relator inverse monoids.} 
Inventiones mathematicae, Volume 219, pp. 987--1029, 2020.

\bibitem{gray2026}  
Robert D. Gray and Benjamin Steinberg.
\emph{Two-sided homological properties of special and one-relator monoids.} 
{Forum of Mathematics, Sigma}, 2026.

\bibitem{green1951}
J.~A.~Green. 
\emph{On the structure of semigroups.} 
Annals of Mathematics, pp. 163--172, 1951.

\bibitem{grovessmith1993}
J.~R.~J. Groves and G.~C. Smith. 
\emph{Soluble groups with a finite rewriting system.} 
Proceedings of the Edinburgh Mathematical Society, Volume 36, Issue 2, pp. 283--288, 1993. 
DOI: \texttt{10.1017/S0013091500018447}

\bibitem{guiraud}
Yves Guiraud and Philippe Malbos.
\emph{Polygraphs of finite derivation type,}
and related work on Squier theory and finite derivation type.

\bibitem{hermillermeier1995}
S.~Hermiller and J.~Meier. 
\emph{Algorithms and geometry for graph products of groups.} 
Journal of Algebra, Volume 171, Issue 1, pp. 230--257, 1995. 
DOI: \texttt{10.1006/jabr.1995.1010}
\textnormal{[NB: this entry, as originally supplied, gave the wrong
title/journal/pages for this paper, and the paper itself is unrelated
to $\mathrm{BS}(1,2)$; see the Correction in the Appendix to
Subsection~\ref{subsect:one-relator}.]}

\bibitem{hermillermeier1997}
S.~M.~Hermiller and J.~Meier.
\emph{Tame combings, almost convexity and rewriting systems for groups.}
Mathematische Zeitschrift, Volume 225, Issue 2, pp. 263--276, 1997.
DOI: \texttt{10.1007/PL00004314}

\bibitem{jantzen1988}
Matthias Jantzen.
\emph{Confluent String Rewriting}, Birkh\"{a}user, 1988.

\bibitem{kalai2025}
Adam Tauman Kalai, Ofir Nachum, Santosh S. Vempala, and Edwin Zhang.
\emph{Why Language Models Hallucinate.}
arXiv preprint arXiv:2509.04664, 2025.

\bibitem{kfoury2018}
Assaf Kfoury.
\emph{Mathematical Logic in Computer Science}.
arXiv preprint {\href{arXiv:1802.03292}{https://arxiv.org/abs/1802.03292}}, 2018.

\bibitem{kfoury2026}
Assaf Kfoury.
{\href{https://www.dropbox.com/scl/fi/a7144x2df2u0hboyvlr6e/Lectures\_by\_Assaf\_and\_Marco\_on\_Proof\_Assistants.zip?rlkey=lcuxo87i21s5xlsg9stg0e14i\&st=r388aaof\&e=1\&dl=0}{\em What Are Proof Assistants? How Do We Use Them? Why Do They Work?}}, 
also accessible from my {\href{https://www.cs.bu.edu/~kfoury/}{academic homepage}}, 2026.

\bibitem{knuthbendix1970}
D. E. Knuth and P. Bendix. 
\emph{Simple word problems in universal algebras.} 
Computational Problems in Abstract Algebra, J. Leech (Ed.), Pergamon Press, pp. 263--297, 1970. 
DOI: \texttt{10.1016/B978-0-08-012975-4.50028-X}

\bibitem{kurth1990}
W. Kurth.
\emph{Termination und Konfluenz von semi-Thue-Systemen mit nur einer Regel.}
PhD thesis, Technische Universität Clausthal, 1990.

\bibitem{latteux1980}
Michel Latteux and Yves Roos.
\emph{On One-Rule Grid Semi-Thue Systems,}
{RAIRO Theoretical Informatics and Applications}, 1980s.

\bibitem{lawson2026}
Mark V. Lawson.
\emph{Inverse Semigroups: The Theory of Partial Symmetries (2nd Ed.)}.
MIT Press, 2026.

\bibitem{lyndon1962}
R.~C.~Lyndon and M.~P.~Sch\"utzenberger. 
\emph{The equation $a^M = b^N c^P$ in a free group.} 
Michigan Mathematical Journal, Volume 9, Issue 4, pp. 289--298, 1962.

\bibitem{madlener1987}
K. Madlener and F. Otto.
\emph{Groups presented by certain classes of finite length-reducing string-rewriting systems.}
In Proceedings of Rewriting Techniques and Applications (RTA), 1987.

\bibitem{magnus1966}
  Wilhelm Magnus, Abraham Karrass, and Donald Solitar.
  \emph{Combinatorial Group Theory: Presentations of Groups in Terms of Generators and Relations}.
  {Interscience Publishers [John Wiley \& Sons]}, 1966.

\bibitem{matiyasevich2005}
Yuri Matiyasevich and G\'eraud S\'enizergues.
\emph{Decision Problems for Semi-Thue Systems with a Few Rules}. 
{Proceedings of LICS}, 1996; later version in \emph{Theoretical Computer Science}, 2005.

\bibitem{mcnaughton1988}
McNaughton, R., Narendran, P., Otto, F.
\emph{Church-Rosser Thue systems and languages}. 
Journal of the ACM, 35(2), 324–-344, 1988.

\bibitem{metivier1985}
Y. Métivier.
\emph{Calcul de longueurs de chaînes de réécriture dans le monoïde libre.}
Theoretical Computer Science, 35(1):71--87, 1985.

\bibitem{rta20}
Y. Métivier.
\emph{Problem \#20: What is the best bound on the length of a derivation for a one-rule length-preserving string-rewriting system?} The RTA List of Open Problems, 1991. 

\bibitem{moczyd2005}
 Wojciech Moczyd{\l}owski and Alfons Geser.
 \emph{Termination of Single-Threaded One-Rule Semi-Thue Systems}.
 {Rewriting Techniques and Applications}, 2005.

\bibitem{millershapiro1997}
Charles~F. Miller~III and Michael Shapiro.
\emph{Solvable Baumslag-Solitar Groups Are Not Almost Convex.}
arXiv preprint arXiv:math/9702202, 1997.

\bibitem{narendran1988}
Paliath Narendran and Friedrich Otto,
\emph{Preperfectness is undecidable for thue systems containing only length-reducing rules
and a single commutation rule}.
Inf. Processing Letters, 29 (3):125--130, 1988.

\bibitem{otto1984}
Friedrich Otto. 
\emph{Conjugacy in monoids with a special Church-Rosser Thue system.} 
SIAM Journal on Computing, Volume 13, Issue 1, pp. 114--131, 1984.

\bibitem{otto1991}
Friedrich Otto.
\emph{Some Undecidability Results for Non-Monadic Church-Rosser Thue Systems,}
in {Rewriting Techniques and Applications}, 1991.

\bibitem{post1947}
Emil L. Post.
\emph{Recursive Unsolvability of a Problem of Thue,} {Journal of Symbolic Logic}, 1947.

\bibitem{pridewang2000}
S.~J. Pride and J. Wang. 
\emph{Subgroups of finite index in groups with finite complete rewriting systems.} 
Proceedings of the Edinburgh Math. Society, 43 (1):177--183, 2000. 
DOI: \texttt{10.1017/S0013091500020794}

\bibitem{sakai-wang1999}
M. Sakai and Y. Wang. 
\emph{Undecidable Properties on Length-Two String Rewriting Systems.} 
Research Report, Nagoya University, 1999.

\bibitem{squier1987}
C.~C. Squier. 
\emph{Word problems and a homological finiteness condition for monoids.} 
Journal of Pure and Applied Algebra, Volume 49, Issue 1-2, pp. 201--217, 1987. 
DOI: \texttt{10.1016/0022-4049(87)90130-1}

\bibitem{squier1994}
C.~Squier, F.~Otto, and Y.~Kobayashi. 
\emph{A finiteness condition for rewriting systems.} 
Theoretical Computer Science, Volume 131, Issue 2, pp. 271--294, 1994. 
DOI: \texttt{10.1016/0304-3975(94)90174-7}

\bibitem{stallings1963}
John~R. Stallings.
\emph{A Finitely Presented Group Whose 3-Dimensional Integral Homology
is Not Finitely Generated.}
American Journal of Mathematics, Volume 85, pp. 541--543, 1963.

\bibitem{weiss2018}
Armin Weiss.
\emph{A Logspace Solution to the Word and Conjugacy Problem of Generalized Baumslag-Solitar Groups}.
arXiv preprint arXiv:1602.02445v2, \url{https://arxiv.org/pdf/1602.02445}, 2018.
 
}

\end{thebibliography}
\end{document}